\documentclass[apj]{emulateapj}
\usepackage{amssymb}
\usepackage{epsfig}
\usepackage{hyperref}

\usepackage{longtable}

\begin{document}


\title{The State of the Warm and Cold Gas in the Extreme Starburst \\at the Core of the Phoenix Galaxy Cluster (SPT-CLJ2344-4243)}


\author{Michael McDonald$^{*\dagger1}$, Mark Swinbank$^2$, Alastair C.\ Edge$^2$, David J.\ Wilner$^3$, Sylvain Veilleux$^{4,5}$, Bradford A.\ Benson$^6$, Michael T.\ Hogan$^2$, Daniel P.\ Marrone$^7$, Brian R.\ McNamara $^8$, Lisa H.\ Wei$^9$, Matthew B.\ Bayliss$^{3,10}$, and Marshall W.\ Bautz$^1$}
\altaffiltext{*}{Email: mcdonald@space.mit.edu}
\altaffiltext{$\dagger$}{Hubble Fellow}
\altaffiltext{1}{Kavli Institute for Astrophysics and Space Research, MIT, Cambridge, MA 02139, USA}
\altaffiltext{2}{Institute for Computational Cosmology, Department of Physics, Durham University, South Road, Durham DH1 3LE}
\altaffiltext{3}{Harvard-Smithsonian Center for Astrophysics, 60 Garden Street, Cambridge, MA 02138, USA}
\altaffiltext{4}{Department of Astronomy, University of Maryland, College Park, MD 20742, USA}
\altaffiltext{5}{Joint Space-Science Institute, University of Maryland, College Park, MD 20742, USA}
\altaffiltext{6}{Kavli Institute for Cosmological Physics, University of Chicago, 5640 South Ellis Avenue, Chicago, IL 60637, USA}
\altaffiltext{7}{Steward Observatory, University of Arizona, 933 North Cherry Avenue, Tucson, AZ 85721}
\altaffiltext{8}{Department of Physics and Astronomy, University of Waterloo, 200 University Ave West, Waterloo, Ontario, Canada}
\altaffiltext{9}{Atmospheric and Environmental Research, 131 Hartwell Avenue, Lexington, MA 02421}
\altaffiltext{10}{Department of Physics, Harvard University, 17 Oxford Street, Cambridge, MA 02138}


\begin{abstract}
We present new optical integral field spectroscopy (Gemini South) and submillimeter spectroscopy (Submillimeter Array) of the central galaxy in the Phoenix cluster (SPT-CLJ2344-4243). This cluster was previously reported to have a massive starburst ($\sim$800 M$_{\odot}$ yr$^{-1}$) in the central, brightest cluster galaxy, most likely fueled by the rapidly-cooling intracluster medium. 
These new data reveal a complex emission-line nebula, extending for $>$30 kpc from the central galaxy, detected at [O~\textsc{ii}]$\lambda\lambda$3726,3729, [O~\textsc{iii}]$\lambda\lambda$4959,5007, H$\beta$, H$\gamma$, H$\delta$, [Ne~\textsc{iii}]$\lambda$3869, and He~\textsc{ii} $\lambda$4686. The total H$\alpha$ luminosity, assuming H$\alpha$/H$\beta$ = 2.85, is L$_{H\alpha}$ = 7.6 $\pm$ 0.4 $\times$10$^{43}$ erg s$^{-1}$, making this the most luminous emission line nebula detected in the center of a cool core cluster. 
Overall, the relative fluxes of the low-ionization lines (e.g., [O~\textsc{ii}], H$\beta$) to the UV continuum are consistent with photoionization by young stars.
In both the center of the galaxy and in a newly-discovered highly-ionized plume to the north of the galaxy, the ionization ratios are consistent with both shocks and AGN photoionization. We speculate that this extended plume may be a galactic wind, driven and partially photoionized by both the starburst and central AGN.
Throughout the cluster we measure elevated high-ionization line ratios (e.g., He~\textsc{ii}/H$\beta$, [O~\textsc{iii}]/H$\beta$), coupled with an overall high velocity width (FWHM $\gtrsim$ 500 km~s$^{-1}$), suggesting that shocks are likely important throughout the ISM of the central galaxy. These shocks are most likely driven by a combination of stellar winds from massive young stars, core-collapse supernovae, and the central AGN.
 In addition to the warm, ionized gas, we detect a substantial amount of cold, molecular gas via the CO(3-2) transition, coincident in position with the galaxy center. We infer a molecular gas mass of M$_{H_2}$ = 2.2 $\pm$ 0.6 $\times$ 10$^{10}$ M$_{\odot}$, which implies that the starburst will consume its fuel in $\sim$30~Myr if it is not replenished. The L$_{IR}$/M$_{H_2}$ that we measure for this cluster is consistent with the starburst limit of 500 L$_{\odot}$/M$_{\odot}$, above which radiation pressure is able to disperse the cold reservoir. The combination of the high level of turbulence in the warm phase and the high L$_{IR}$/M$_{H_2}$ ratio suggests that this violent starburst may be in the process of quenching itself. We propose that phases of rapid star formation may be common in the cores of galaxy clusters, but so short-lived that their signatures are quickly erased, and appear only in a subsample of the most strongly cooling clusters.

\end{abstract}




\section{Introduction}

The Phoenix cluster \citep[SPT-CLJ2344-4243;][]{mcdonald12c}, which was discovered with the South Pole Telescope \citep[SPT;][]{carlstrom11} and initially reported by \cite{williamson11}, is, at $z=0.597$, the most X-ray luminous galaxy cluster yet discovered \citep[L$_{\rm{2-10keV}}=8.2\times10^{45}$ erg s$^{-1}$;][]{mcdonald12c}. This exceptionally high X-ray luminosity is due to the combination of a very massive galaxy cluster \citep[M$_{500} = 12.6 \times 10^{14}$ M$_{\odot}$;][]{mcdonald12c}, a heavily-obscured central AGN \citep[L$_{\rm{2-10keV, unabsorbed}}=3\times10^{45}$ erg s$^{-1}$;][]{mcdonald12c}, and an extreme cooling flow \citep[\.{M}$_{\rm{classical}} \sim2000$ M$_{\odot}$ yr$^{-1}$;][]{mcdonald13b}. 
Unlike nearby ``cool core clusters'' -- characterized by dense, cool central regions -- which typically convert only a few percent of the cooling intracluster medium (ICM) into stars \citep[e.g.,][]{johnstone87,mcnamara89,odea08,mcdonald11b}, the central galaxy in the Phoenix cluster (hereafter Phoenix A) appears to be experiencing an $\sim$800 M$_{\odot}$ yr$^{-1}$ starburst \citep{mcdonald13a} -- consuming roughly 30--40\% of the expected cooling flow. This estimate of the star formation rate in Phoenix A, which is based on \emph{Hubble Space Telescope} far-ultraviolet imaging, is in line with the expected efficiency of star formation in giant molecular clouds \citep[$\sim$20--50\%;][]{kroupa01a,lada03}, suggesting that the ``cooling flow problem'' \citep[for a review, see][]{fabian94} may not be as severe in this unique system.

Morphologically complex, extended nebulae of warm (10$^4$~K), ionized gas are nearly ubiquitous in cool core clusters like the Phoenix cluster \citep{hu85, johnstone87, heckman89, crawford99, edwards07, hatch07, mcdonald10, mcdonald11a} -- so much so that H$\alpha$ luminosity has often been used as an alternative classification of rapid ICM cooling \citep{donahue92, samuele11, mcdonald11c}. The most spectacular such nebulae is found in the nearby Perseus cluster \citep{conselice01, fabian03,hatch06}, with multiple filaments extending radially from the central galaxy to the cooling radius ($\sim$60~kpc). While it is generally assumed that this warm gas has cooled from the ICM, it has become clear that there is probably not a single ionization source responsible for all of the optical line emission observed in cluster cores. Instead, photoionization from young stars \citep[e.g.,][]{johnstone87, mcnamara89, allen95, crawford99, hatch07, mcdonald12a}, slow shocks \citep[e.g.,][]{farage10,mcdonald12a}, condensing intracluster gas \citep[e.g.,][]{voit90, donahue91, voit94}, and particle heating \citep[e.g.,][]{ferland09, fabian11} likely contribute at levels varying from cluster to cluster, and can even vary spatially within a given cluster.

In addition to warm, ionized gas, cool core clusters tend to have massive reservoirs of cold molecular gas \citep{edge01, edge02, salome03}. This cold gas reservoir is typically centrally-concentrated about the brightest cluster galaxy (BCG), but has also been found coincident with extended H$\alpha$ emission \citep[e.g.,][]{edge03, salome04, hatch05, salome08, salome11, mcdonald12b}. These cold reservoirs have masses of order 10$^{9-11.5}$ M$_{\odot}$ \citep[][]{edge01} and are typically interpreted as the final stage of the residual cooling flow, with some feedback mechanism \citep[e.g., AGN: see recent reviews by][]{fabian12,mcnamara12} preventing $\sim$90\% of the cooling ICM from reaching this cold state.

Here we present new observations of both the warm (Gemini Multi-Object Spectroscopy) and cold (Sub-millimeter Array) gas in the core of the Phoenix cluster. With these data, we will attempt to understand the mass, distribution, kinematics, and ionization state of the different gas phases in this most extreme system. Coupled with the existing X-ray (\emph{Chandra X-ray Observatory}), optical (\emph{Hubble Space Telescope}), and infrared (\emph{Herschel Space Observatory}) data, we will attempt to build a more complete picture of the massive, cooling flow-fueled starburst in the central galaxy of the Phoenix cluster. Throughout this paper, we assume H$_0$~=~70 km s$^{-1}$ Mpc$^{-1}$, $\Omega_M$ = 0.27, and $\Omega_{\Lambda}$ = 0.73.


\section{Data Reduction \& Analysis}
Below we describe the acquisition, reduction, and analysis of data from both the Gemini-South Multi-Object Spectrograph (GMOS-S) and the Sub-millimeter Array (SMA). Data from these observatories were acquired via Director's Discretionary Time proposals DT-2012B-002 and 2013A-S070, respectively.

\subsection{GMOS-S Data Reduction}

Spectro-imaging observations of Phoenix A were obtained with
the GMOS-S IFU 2012 November 19 and 2012 November 16 during dark time
with photometric conditions and $\sim $\,0.6$''$ $V$-band seeing
(corresponding to a physical resolution of $\sim $\,4.0\,kpc at $z=0.597$, the
redshift of our target).  The observations employed the single slit
mode, resulting in a field of view of 5$''\times$7$''$.  We used the
R400 filter centered at 8626\AA\ which provides wavelength coverage
from 5360--9600\AA\ at a spectral resolution of
$\lambda/\Delta\lambda\sim $\,2000 (3.9\AA\ per pixel).  To cover the
galaxy, we used a small mosaic: a central pointing plus one
pointing either side (separated by a full IFU width).  The final field
of view is therefore 9$^{\prime\prime}$$\times$\,5$^{\prime\prime}$.  The exposure time for the
central pointing was 2.4\,ks (split into two 1.2\,ks exposures) whilst
the two outer pointings each have 4.8\,ks (split into four 1.2\,ks
exposures).  

\begin{figure}[htb]
\centering
\includegraphics[width=0.49\textwidth]{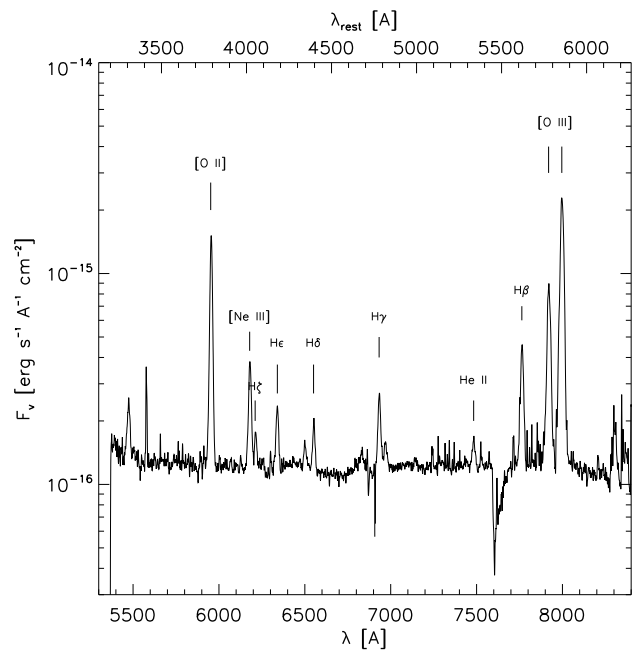}
\caption{Collapsed 1-D spectrum using the full GMOS field of view. The strong absorption lines at $\sim$6800\AA\ and $\sim$7800\AA\ are due to O$_2$ in the Earth's atmosphere. This figure confirms the flat spectral energy distribution presented in \cite{mcdonald12c}, as well as the presence of both high- and low-ionization emission lines.
}
\label{fig:fullspec}
\end{figure}

To reduce the data, we used the GMOS data reduction pipeline to extract
and wavelength-calibrate the spectra of each IFU element.  Variations
in fiber-to-fiber response were removed using twilight flats, and the
wavelength calibration employed a CuAr arc lamp.  However, since the arc
lamps were taken several days after the science observations, we applied a
small zero-point correction to the wavelength solution in each fiber for each
science exposure using the sky OH emission.   Flux calibration was carried out 
using observations of the standard star LTT3218 using the same
observational setup as the science observations. The collapsed 1-D spectrum of the full flux-calibrated data cube is shown in Figure \ref{fig:fullspec}.

\begin{figure}[htb]
\centering
\includegraphics[width=0.4\textwidth,trim=30.2cm 0cm 0cm 0cm,clip=true]{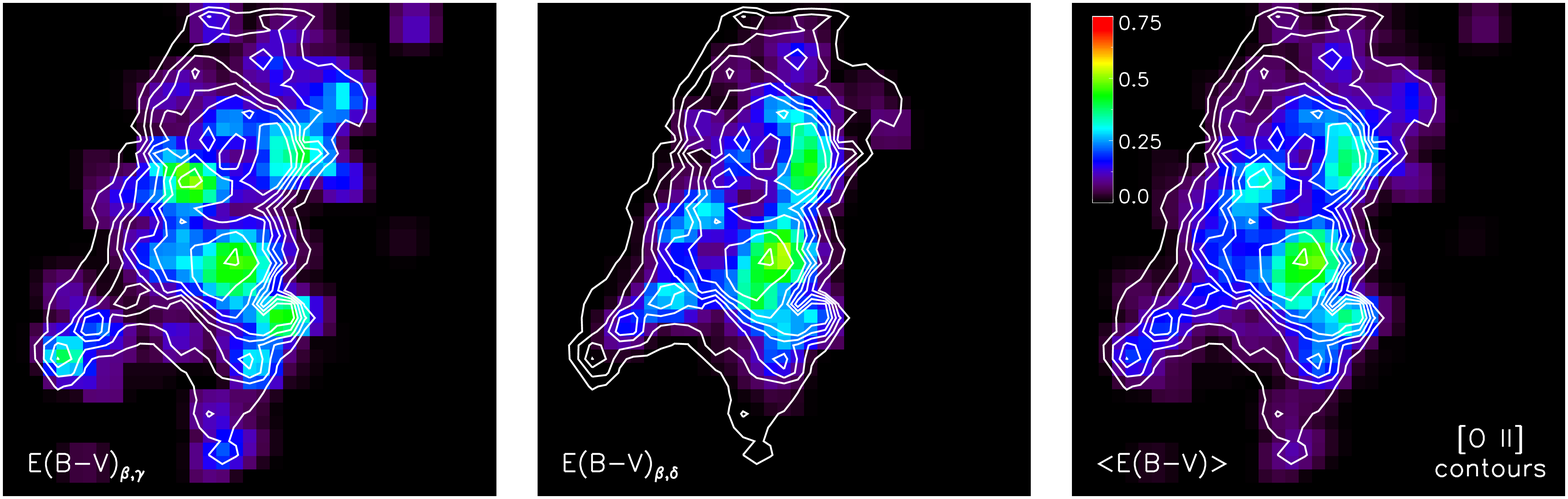}
\caption{
%
Combined reddening map, derived using the H$\beta$, H$\gamma$, and H$\delta$ emission lines. White contours show where the reddening-corrected [O~\textsc{ii}]$\lambda$3727 emission lies, for comparison. We note the strong resemblance, both in morphology and absolute value, to the reddening map derived in \cite{mcdonald13a}, which was based on the UV slope. 
}
\label{fig:ebv}
\end{figure}


\subsection{GMOS-S Data Analysis}
Each spatial pixel in the GMOS-S IFU cube corresponds to an independent spectrum covering 5360--9600\AA\ (rest-frame 3356--6011\AA). These single-pixel spectra were first fit with a single stellar population (SSP) in order to model the continuum emission and absorption lines. SSPs were generated using the publicly-available Sed@ code\footnote{\tt{\url{http://www.iaa.es/~mcs/sed@/}}} and made available by \cite{gonzalez-delgado05}. These models cover ranges of 0.001--0.04 in metallicity and 10$^6$--10$^{10}$ yr in age, and we apply a range of reddening models spanning 0 $<$ E(B-V) $<$ 3. We choose the combination of age, metallicity, and reddening that minimizes $\chi^2$ over the full spectrum, with the normalization being a free parameter. 
The best-fit continuum spectrum was subtracted from each spectrum, leaving only emission lines. We note that this part of the analysis was performed in order to ensure accurate Balmer line fluxes (factoring in stellar absorption to the estimates of emission). However, we find that, given the strong Balmer emission in this system, we obtain very similar results whether we assume no Balmer absorption or perform a more complicated continuum modeling as described above.

Each of the resulting continuum-subtracted spectra was smoothed in wavelength space, with a smoothing scale of 8 spectral pixels, in order to improve the quality of emission-line fits. We performed simultaneous fits of the [O~\textsc{ii}]$\lambda\lambda$3726,3729, [O~\textsc{iii}]$\lambda\lambda$4959,5007, H$\beta$, H$\gamma$, H$\delta$, [Ne~\textsc{iii}]$\lambda$3869, and He~\textsc{ii}~$\lambda$4686 emission lines, requiring all emission lines to have identical redshifts and linewidths. This resulted in maps of emission line fluxes and radial velocities as a function of position. Finally, we went back and fit the unsmoothed [O~\textsc{iii}]$\lambda\lambda$4959,5007 lines, using the previously-measured redshift and flux as priors, in order to determine the gas kinematics as a function of position. During this iteration, we allowed multiple kinematic components for each line. Measured velocity dispersions were corrected for the instrumental linewidth of 3.1\AA\ at the wavelength of the [O~\textsc{iii}]$\lambda$5007 emission line.

The H$\beta$, H$\gamma$, and H$\delta$ lines were used to estimate the amount of reddening at each position. We assume case B recombination, using intrinsic values of H$\gamma$/H$\beta=0.47$ and H$\delta$/H$\beta=0.26$ \citep{osterbrock89}. The measured Balmer ratios, combined with the \cite{cardelli89} extinction curve, yield two independent maps of E(B-V), which have qualitatively similar magnitudes and morphologies. These two maps are averaged. The resulting map of E(B-V), shown in Figure \ref{fig:ebv}, is used to correct all measured fluxes for intrinsic reddening, assuming a dust-screen model.

\begin{figure}[htb]
\centering
\includegraphics[width=0.49\textwidth]{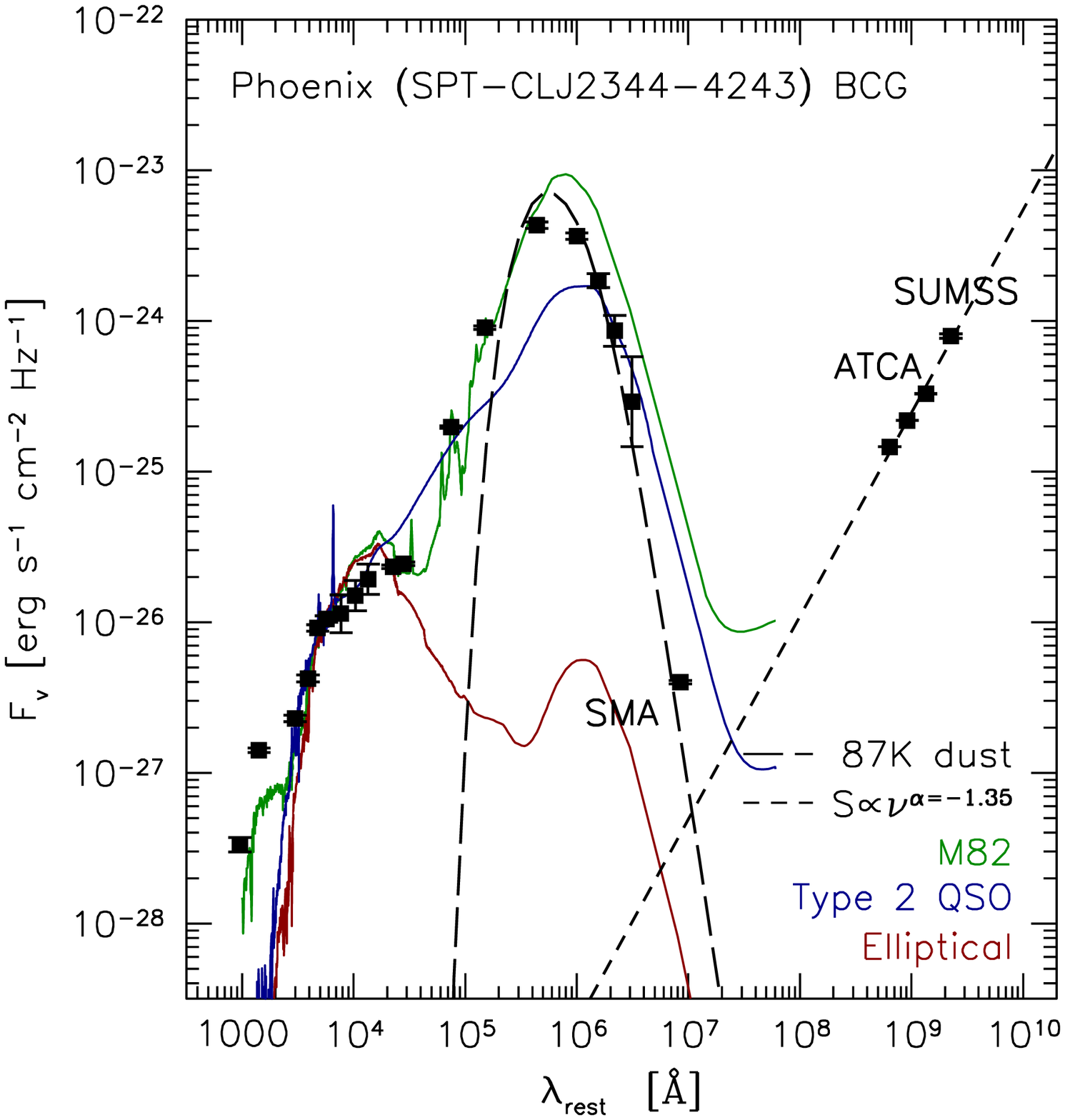}
\caption{Spectral energy distribution (SED) for the central galaxy in the Phoenix cluster, Phoenix A. The UV, optical, and infrared data ($\lambda<10^7$\AA) were taken from \cite{mcdonald12c}. We show template SEDs from \cite{polletta07} for comparison, demonstrating that the UV-optical-IR SED most resembles a dusty, star-forming galaxy (e.g., M82). The two right-most data points are from archival SUMSS \citep[833 MHz;][]{mauch03} and ATCA (2.1 GHz; PI R.\ Kale) observations, while the point at $\sim10^7$\AA\ is from this work (220.7 GHz). Extrapolating the radio data (assuming synchrotron emission) and the far-IR data \citep[assuming a 47K blackbody;][]{mcdonald12c}, we find that the continuum emission in our SMA observations is most likely dominated by cold dust emission from star-forming regions, rather than synchrotron emission from the radio galaxy.
}
\label{fig:sed}
\end{figure}

\subsection{SMA Data Reduction and Analysis}

We used the SMA\footnote{The Submillimeter Array is a joint project between
the Smithsonian Astrophysical Observatory and the Academica Sinica Institute
of Astronomy and Astrophysics and is funded by the Smithsonian Institution
and the Academica Sinica.} \citep{ho04} on Mauna Kea, Hawaii, to observe
the core of the Phoenix cluster on 2013 Aug 13 and 2013 Aug 18. The five available
array antennas were arranged in a compact configuration that gave baseline
lengths from 6 to 68 meters. The weather conditions were very good to
excellent on these two dates, with $\tau_{\rm 225~GHz}$ of 0.10 and 0.05
measured by the atmospheric opacity monitor located at the nearby Caltech
Submillimeter Observatory. The receivers were tuned to search for emission
from the CO J=3-2 line (rest frequency 345.796 GHz) at the cluster redshift
of $z=0.597$. The correlator was configured to process two IF bands, each of
width 1.968 GHz, centered at $\pm5$ GHz and at $\pm7$ GHz from an LO frequency
of 221.435 GHz. This setup resulted in velocity coverage of approximately
$-1240$ to $4270$~km~s$^{-1}$ around $z=0.597$ for the CO J=3-2 line in the LSB (aside from
a small gap of about 32 MHz between the two IF bands). The channel spacing
was 0.8125 MHz, corresponding to 1.13~km~s$^{-1}$. The pointing center was
chosen to be at the center of the BCG, $\alpha = 23^h44^m43\fs96$,
$\delta = -42\degr43\arcmin12\farcs2$ (J2000), and the $\sim58''$ (FWHM) field
of view was set by the primary beam size of the individual array antennas.
The basic observing loop comprised 2.5 minutes on J2258-279,
2.5 minutes on J2248-325, and 7.5 minutes on the Phoenix.  The absolute flux
scale was set with an accuracy of $\sim10\%$ using observations of Uranus at
the start of each track. The passband shape was determined using observations
of 3C84 at the end of each track. Time dependent complex gains were derived
from the frequent observations of J2258-279 (0.88 Jy), and the efficacy of
the solutions were verified by application to J2248-325 (0.18 Jy). All of the
calibration steps were performed with the IDL based {\tt MIR} software.
Imaging and deconvolution were done in {\tt MIRIAD} with standard routines.
The synthesized beam size with natural weight was $7\farcs5 \times 2\farcs5$ (51$\times$17 kpc),
position angle $-6$ degrees; the substantial north-south elongation of the
beam results from the low declination of the cluster. After combining the two
tracks, the rms noise was 4~mJy~beam$^{-1}$ in 200~km~s$^{-1}$ velocity bins.

\begin{figure*}[htb]
\centering
\includegraphics[width=0.99\textwidth]{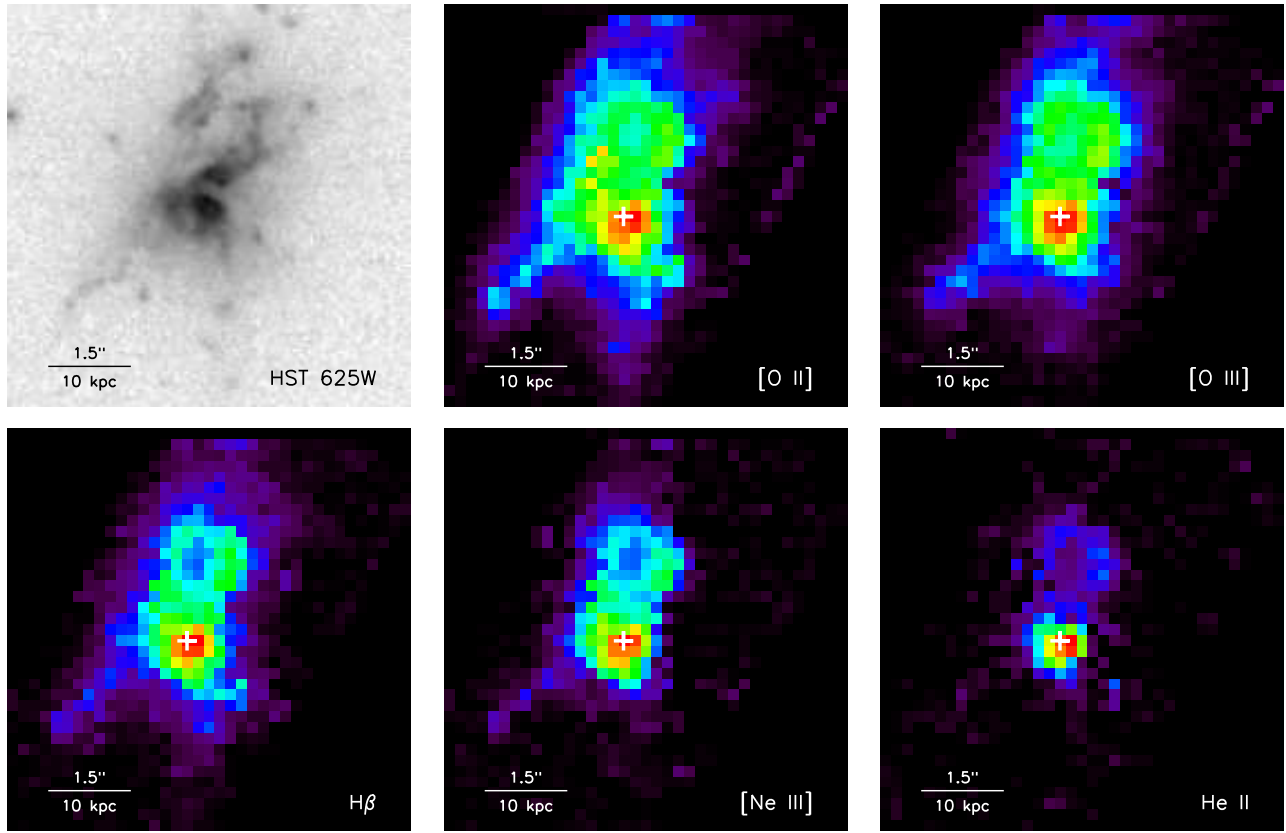}
\caption{Individual reddening-corrected emission line maps for [O~\textsc{ii}], [O~\textsc{iii}], H$\beta$, [Ne~\textsc{iii}], and He~\textsc{ii}. In the upper left corner, we show an HST image in the F625W band from \cite{mcdonald13a}. The peak of the reddening-corrected continuum emission (presumably the central AGN) is denoted with a white cross in all panels. The loop of UV emission to the north of the central galaxy is visible in all 5 emission lines, along with the thin, extended filaments to the southeast and southwest. The presence of extended He~\textsc{ii} suggests a very hard ionization source, possibly an AGN-driven outflow or an exceptionally massive young stellar population (i.e., Wolf-Rayet stars).}
\label{fig:lines}
\end{figure*}

In the continuum map (combined USB and LSB, effective frequency of 220.7 GHz) we detect a point source at $\alpha = 23^h44^m43\fs89$, $\delta = -42\degr43\arcmin12\farcs29$ (J2000), with a flux of 0.25 $\pm$ 0.03 mJy, compared to 79.2 mJy at 833 MHz from the SUMSS database \citep{mauch03}. In order to determine whether the SMA continuum flux is due to star formation or synchrotron emission, we consider archival ATCA 1.4, 2.0, and 2.9 GHz data (PI: R.\ Kale). These data, when combined with the aforementioned SUMSS data, suggests a low-frequency radio slope of $\alpha \sim -1.35$. At a frequency of 220 GHz, this would result in a flux of 0.04 mJy, or a factor of $\sim$6 lower than what we measure. On the contrary, if we extrapolate the best-fit blackbody spectrum to the far-infrared data, we intersect the SMA continuum emission almost exactly (see Figure \ref{fig:sed}). Thus, we conclude that the SMA continuum emission is not probing the radio-loud AGN, and so we will not consider it further in this work.

\section{Results}

Below, we summarize the results of this study, first based on the optical GMOS-IFU data tracing the warm (10$^4$K) gas, followed by our SMA CO(3-2) observations tracing the cold ($\sim$10$^2$K) gas. For the most part, we will defer discussing the implications of these results to \S4.

\subsection{Optical Emission Line Maps}
In order to achieve maximum signal-to-noise, we heavily binned and smoothed each spectrum before measuring the line flux. As a result, we measure the total emission line fluxes, even in cases where there are multiple velocity components. In Figure \ref{fig:lines} we show emission line maps for the [O~\textsc{ii}]$\lambda\lambda$3726,3729, [O~\textsc{iii}]$\lambda\lambda$4959,5007, H$\beta$, H$\gamma$, H$\delta$, [Ne~\textsc{iii}]$\lambda$3869, and He~\textsc{ii}~$\lambda$4686 emission lines, compared to a rest-frame blue image. Overall, the [O~\textsc{ii}] emission has a very similar morphology to the blue continuum, suggesting that this gas may be photoionized by young stars. The presence of extended, filamentary [O~\textsc{ii}] extending $>$30~kpc from the center of Phoenix A is reminiscent of the complex H$\alpha$ filaments in nearby systems such as NGC~1275 \citep[e.g.,]{conselice01, fabian03, hatch06} and Abell~1795 \citep{cowie83,mcdonald09}. Luminous, extended [O~\textsc{iii}] emission is not, however, typically found in the central galaxies of cool core clusters. The additional presence of extended, high-ionization lines such as [Ne~\textsc{iii}] and He~\textsc{ii} suggest that an additional, harder, ionization source may also be operating on large physical scales in this system. All five emission lines, along with the stellar continuum, share a common peak. This nucleus is exceptionally bright in He~\textsc{ii}, suggesting that it is most likely ionized by the powerful central AGN \citep{mcdonald12c,ueda13}.

We quote, in Table \ref{table:flux}, the integrated fluxes of all five aforementioned emission lines, both in the central, nuclear region (0.6$^{\prime\prime}$$\times$0.6$^{\prime\prime}$ box), as well as in our total GMOS-IFU field of view.
The total, extinction-corrected [O~\textsc{ii}] flux in this system is f$_{[O II]}$ = 5.0 $\pm$ 0.3 $\times$ 10$^{-14}$ erg s$^{-1}$ cm$^{-2}$, which corresponds to a luminosity of L$_{[O II]}$ = 7.6 $\pm$ 0.5 $\times$ 10$^{43}$ erg s$^{-1}$. Following \cite{kewley04}, the [O~\textsc{ii}]-derived star formation rate, excluding the nuclear contribution, is 431 $\pm$ 26 M$_{\odot}$ yr$^{-1}$, while following \cite{kennicutt98} predicts a higher value of 918 $\pm$ 55 M$_{\odot}$ yr$^{-1}$. Our estimate based on the UV luminosity \citep{mcdonald13a} lies between these two values, at 798 $\pm$ 42 M$_{\odot}$ yr$^{-1}$.

\begin{center}
\LTcapwidth=0.49\textwidth
\begin{longtable}{c c c}
\hline\hline
Line & f$_{nucleus}$ & f$_{total}$ \\
 & [10$^{-15}$ erg s$^{-1}$ cm$^{-2}$] & [10$^{-14}$ erg s$^{-1}$ cm$^{-2}$] \\
 \hline
H$\beta$ & 2.90 $\pm$ 0.14 & 1.76 $\pm$ 0.10\\
\textrm{[O~\textsc{ii}]} &  7.12 $\pm$ 0.36& 5.04 $\pm$ 0.26\\
\textrm{[O~\textsc{iii}]} & 30.4 $\pm$ 1.53 & 12.9 $\pm$ 0.65\\
\textrm{[Ne~\textsc{iii}]} & 2.77 $\pm$ 0.15 & 1.30 $\pm$ 0.08\\
\textrm{He~\textsc{ii}} & 0.47 $\pm$ 0.03 & 0.22 $\pm$ 0.02\\
\hline
\\
\caption{Reddening-corrected fluxes for various emission lines based on the GMOS-IFU data presented here. The ``nucleus'' corresponds to a 3$\times$3 pixel region centered on the [O\textsc{iii}] peak.}
\label{table:flux}
\end{longtable}
\end{center}

\subsection{Optical Emission Line Ratios}

\begin{figure*}[htb]
\centering
\includegraphics[width=0.98\textwidth]{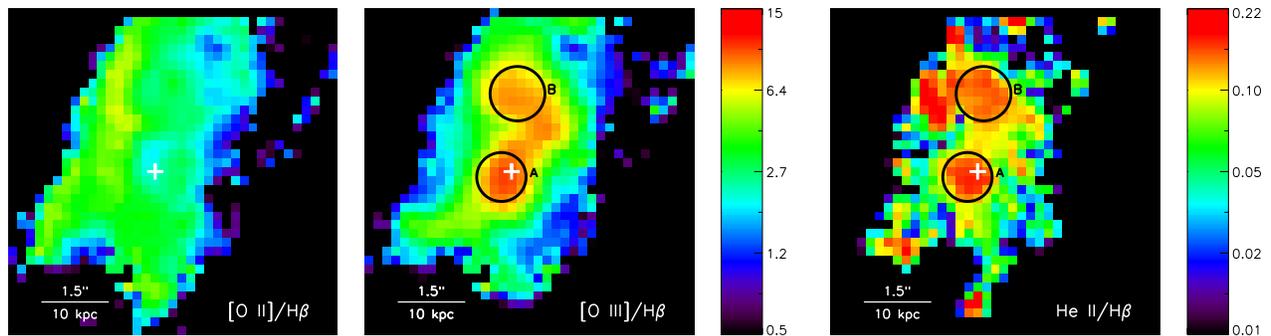}
\caption{Reddening-corrected emission line ratio maps for the [O~\textsc{iii}]/H$\beta$,  [O~\textsc{ii}]/H$\beta$, and He~\textsc{ii}/H$\beta$ ratios. Both the [O~\textsc{iii}]/H$\beta$ and He~\textsc{ii}/H$\beta$ ratios show an extended region of highly-ionized material to the north of the nucleus, potentially due to an AGN-driven outflow. Overall, the [O~\textsc{ii}]/H$\beta$ is smooth, with a mean value around [O~\textsc{ii}]/H$\beta$ $\sim$ 2. The fact that the extended plume of highly-ionized gas is not visible in the [O~\textsc{ii}]/H$\beta$ map suggests that there are likely two (or more) important ionization sources in this system. We have identified the two peaks in the He~\textsc{ii}/H$\beta$ map as region A and B, which will aid the discussion of these regions throughout the remainder of the paper.}
\label{fig:lineratims}
\end{figure*}

\begin{figure*}[p]
\centering
\includegraphics[width=0.95\textwidth]{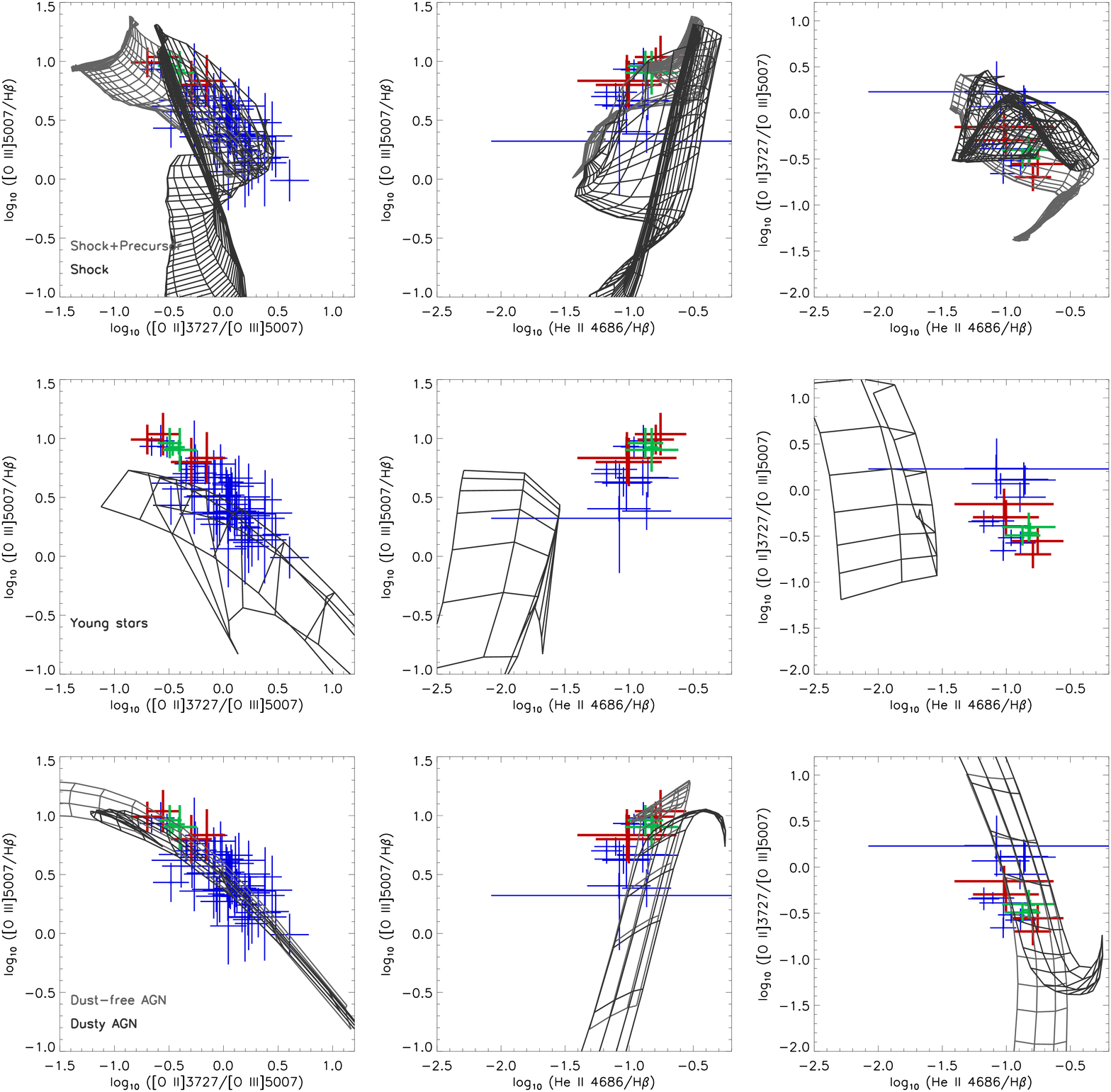}
\caption{Reddening-corrected diagnostic line-ratio plots, motivated by \cite{groves04b}. Model expectations are overplotted for shocks \citep[upper row;][]{allen08}, stellar photoionization \citep[middle row;][]{kewley01}, and AGN photoionization \citep[bottom row;][]{groves04b}. In all panels, red and green points correspond to ``Region A'' and ``Region B'' from Figure \ref{fig:lineratims}, respectively, while the remaining points are shown in blue. Each data point corresponds to a 3$\times$3 pixel area, in order to improve signal-to-noise. This figure demonstrates that the AGN and shock models perform equally well at reproducing all of the observed line ratios, while stellar photoionization is able to adequately describe the data only in the faint, low-ionization outskirts, where the [O~\textsc{iii}]/H$\beta$ ratio is low ($\lesssim$3) and we do not detect He~\textsc{ii}.}
\label{fig:linerats}
\end{figure*}

\begin{figure*}[htb]
\centering
\includegraphics[width=0.9\textwidth]{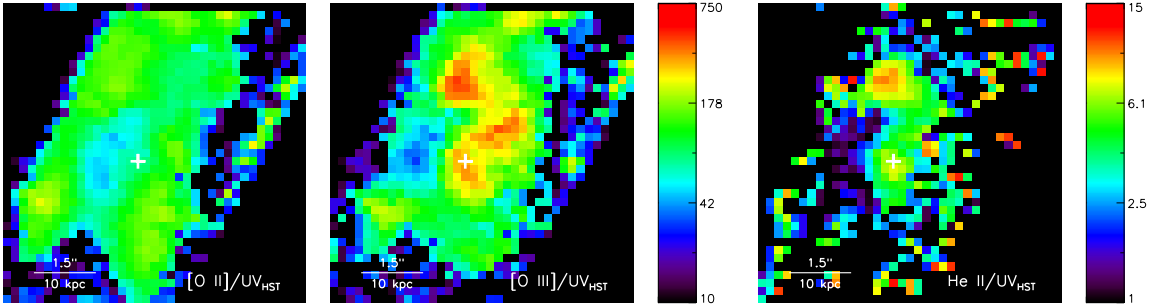}
\caption{Ratio maps of various emission lines and the near-UV continuum level. In all panels, we use the HST F475W image for the near-UV, which corresponds to rest-frame 3000\AA. Similar to figure \ref{fig:lineratims}, the [O~\textsc{iii}]/UV map reveals a highly-ionized plume extending to the north of the galaxy center. This plume is not visible in the [O~\textsc{ii}]/UV map, which is overall very smooth and consistent with the ratio expected for star-forming regions \citep{kennicutt98}. This figure suggests that the low-ionization [O~\textsc{ii}] line has a different ionization source than the high-ionization [O~\textsc{iii}] and He~\textsc{ii} lines.}
\label{fig:uvrats}
\end{figure*}

\begin{figure*}[htb]
\centering
\includegraphics[width=0.9\textwidth]{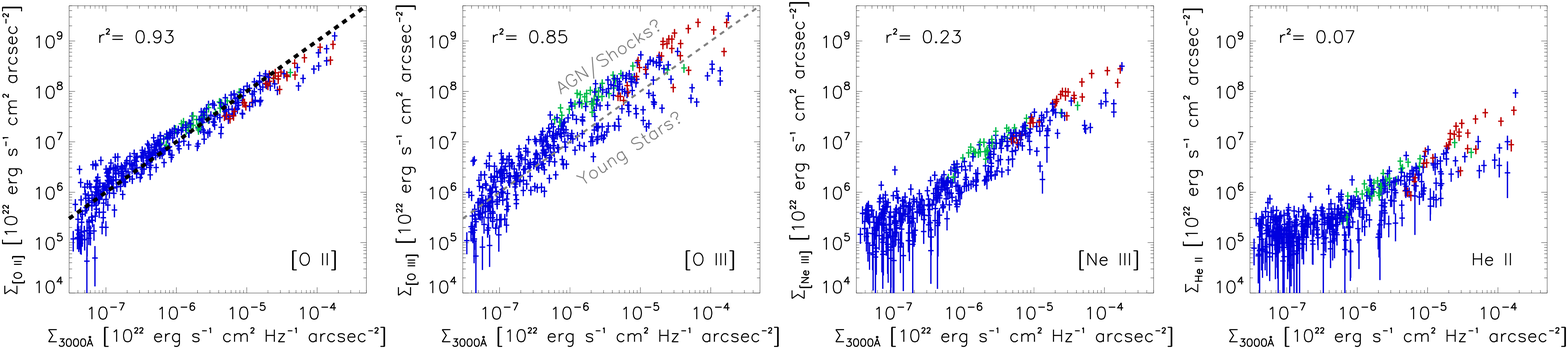}\\
\caption{Pixel-to-pixel emission line surface brightness as a function of near-UV surface brightness. The dotted line in the left-most panel corresponds to the expectation from \cite{kennicutt98}, while the point color follows the same scheme as Figure \ref{fig:linerats}. In the upper left of each panel we provide the Pearson correlation coefficient (r$^2$). This figure demonstrates that the [O~\textsc{ii}]/UV ratio is well-modeled by stellar photoionization, while the higher-ionization lines seem to require an ionization source unrelated to the underlying UV continuum. Interestingly, the [O~\textsc{iii}] vs UV relation appears to fork, with two tracks separated by nearly an order of magnitude, suggesting the presence of two competing ionization sources. We have highlighted this scenario by showing the separation with a dashed gray line and labeling the ``AGN/shocks'' and ``young stars'' tracks.}
\label{fig:uvoii}
\end{figure*}

The ratios of optical emission lines such as [N~\textsc{ii}]/H$\alpha$, [S~\textsc{ii}]/H$\alpha$, and [O~\textsc{i}]/H$\alpha$ are commonly used to separate low-ionization processes such as stellar photoionization from high-ionization processes such as fast shocks \citep[e.g.,][]{baldwin81, veilleux87, kewley06}. Such diagnostics have been applied to the complex emission-line nebulae in cool core clusters \citep[e.g.,][]{voit97, crawford99, hatch06, mcdonald12a}, finding LINER-like line ratios which could be due to a combination of photoionization from young stars and slow shocks \citep{mcdonald12a}. Due to its higher redshift, we are unable to perform these diagnostics for Phoenix A, as most of the relevant lines are redshifted to the near-infrared. Instead, we consider the ratios of [O~\textsc{iii}]/H$\beta$,  [O~\textsc{ii}]/H$\beta$, and He~\textsc{ii}/H$\beta$ in Figure \ref{fig:lineratims}. These maps show a plume of material to the north-west of the galaxy center with very high ionization ([O~\textsc{iii}]/H$\beta$ $>$ 5, He~\textsc{ii}/H$\beta$ $>$ 0.1). This high-ionization material is not seen in the [O~\textsc{ii}]/H$\beta$ map, suggesting that the [O~\textsc{ii}] and [O~\textsc{iii}] have different origins. Outside of the central region, the [O~\textsc{iii}]/H$\beta$ emission-line ratio drops to roughly unity, which is typical for young star-forming regions.

In Figure \ref{fig:linerats}, we compare the measured optical line ratios in 3$\times$3 pixel regions to models of fast radiative shocks \citep{allen08}, photoionization from young stars \citep{kewley01}, and photoionization from AGN \citep{groves04b}, based on three blue emission-line ratio diagnostic plots from \cite{groves04b}. We find that much of the high-ionization emission is inconsistent with stellar photoionization, and fully consistent with both shocks and AGN. In the outer region of the galaxy, where we are unable to detect [He~\textsc{ii}], the emission is consistent with all three models, including photoionization. Given the presence of a highly-luminous type-2 QSO in the center of this galaxy \citep{mcdonald12c, ueda13}, it is reasonable to conclude that the warm gas in the central $\sim$5~kpc (red points) is photoionized by the AGN. The plume of high-ionization gas extending $\sim$15~kpc to the north of the nucleus (green points), may instead be heated by shocks. We will return to this discussion in the next section, where we will incorporate the kinematics into these diagnostics.

Overall, the line ratio diagnostics presented in Figure \ref{fig:lineratims} and \ref{fig:linerats} suggest that there may be two to three separate ionization mechanisms at work in Phoenix A, resulting in a localized peak at the galaxy nucleus, a highly-ionized plume of material to the north of the galaxy center, and a more extended, low-ionization component with the same overall morphology as the bright UV emission \citep{mcdonald13a}.

Utilizing our deep near-UV imaging \citep{mcdonald13a}, we can also do pixel-by-pixel comparisons of the emission line fluxes to the UV fluxes. The results of such an experiment are presented in Figure \ref{fig:uvrats}, where we compare the [O~\textsc{ii}], [O~\textsc{iii}], and He~\textsc{ii} fluxes to the near-UV continuum ($\sim$3000\AA) from HST data. These maps, once again, reveal a plume of highly-ionized material to the north of the galaxy center, with high [O~\textsc{iii}]/UV and He~\textsc{ii}/UV ratios which are inconsistent with star-forming regions. In contrast, the [O~\textsc{ii}]/UV ratio map is fairly smooth, with a mean level consistent with the expectation for a star-forming region \citep[e.g.,][]{kennicutt98, kewley04}. This figure suggests that the low-ionization and high-ionization lines may have different origins, particularly along the plume of warm gas extending north of the galaxy center.

In Figure \ref{fig:uvoii} we compare the pixel-to-pixel emission line surface brightnesses to the cospatial rest-frame far-UV surface brightness. We confirm that the [O~\textsc{ii}] flux is strongly correlated with the UV surface brightness, with a ratio consistent with the expectations from \citep{kennicutt98}. The downturn in [O~\textsc{ii}] at low UV surface brightness is interpreted as an age gradient -- older stars still produce significant UV luminosity but their spectra are not hard enough to ionize the warm gas. 
Comparing the pixel-to-pixel fluxes in various emission lines to the UV continuum, we find the strongest correlation between [O~\textsc{ii}] and the UV surface brightness (r$^2$ = 0.93), with the high-ionization lines being only weakly correlated with the UV surface brightness (r$^2$ = 0.23 and 0.07, for [Ne~\textsc{iii}] and He~\textsc{ii}, respectively). There appears to be two separate sequences in the [O~\textsc{iii}] vs UV plot, again suggesting two (or more) sources of ionization -- one which produces Kennicutt-like ratios (lower [O~\textsc{iii}]/UV) and one which appears over-ionized (high [O~\textsc{iii}]/UV). This plot further supports the emerging picture thus far that the low-ionization lines (e.g., H$\beta$, [O~\textsc{ii}]) are produced in star-forming regions, while the high-ionization lines (e.g., [O~\textsc{iii}], [Ne~\textsc{iii}], He~\textsc{ii}) are produced by a secondary process (e.g., shocks, AGN photoionization) which is uncorrelated with the local UV background.

\subsection{Warm (10$^4$K) Gas Kinematics}

\begin{figure}[htb]
\centering
\includegraphics[width=0.49\textwidth]{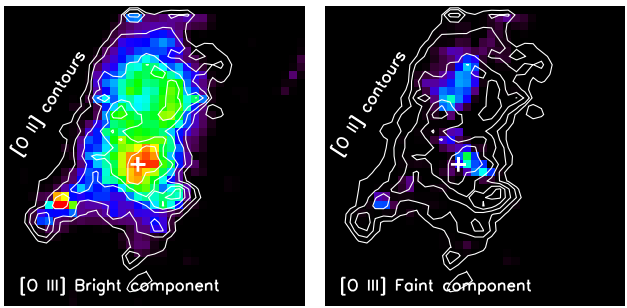}
\caption{[O~\textsc{iii}]$\lambda$5007 emission line maps for the bright and faint components when the two distinct kinematic components are allowed in the spectral fitting. The physical scale here is the same as in Figure \ref{fig:lines}. This figure demonstates that there is significant velocity structure both in the nucleus and along the northern plume, identified in Figures \ref{fig:lineratims} and \ref{fig:uvrats}.}
\label{fig:doubleline}
\end{figure}

The measurement of line redshifts and widths is complicated by the fact that, in a significant number of spectra, we find multiple velocity components. We address this by attempting to fit two distinct velocity components to the [O~\textsc{iii}]$\lambda\lambda$4959,5007 doublet without any smoothing/binning of the spectra. We show, in Figure \ref{fig:doubleline}, the relative intensities in regions where we detect emission line splitting. We find that the region to the northeast of the galaxy center, which we showed previously to have bright [O~\textsc{iii}] and He~\textsc{ii} emission relative to the H$\beta$ and UV emission, has a substantial contribution to its flux from two kinematically distinct components. This suggests a very dynamic environment, as we will discuss in more detail in \S4.

\begin{figure*}[htb]
\centering
\includegraphics[width=0.99\textwidth]{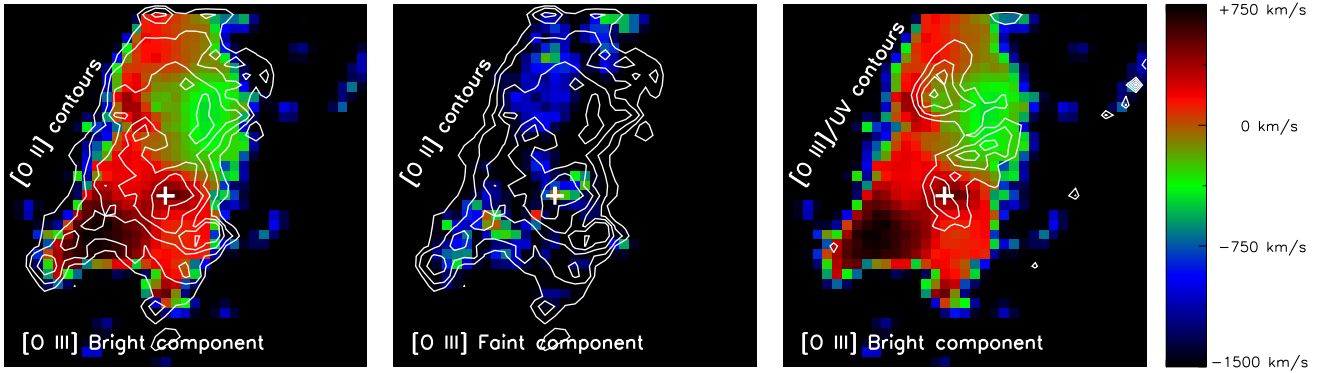}
\caption{Velocity maps of the warm ionized gas in the core of the Phoenix cluster. All velocities are relative to $z=0.597$, the systemic velocity of the central galaxy. The physical scale in all panels is identical to Figure \ref{fig:lines}.  In the central panel we show the line-of-sight velocity of the second, fainter component in cases where multiple velocity components were detected (see also Fig.\ \ref{fig:doubleline}). In the central and left panels, the white contours represent the [O~\textsc{ii}] emission, while the right-most panel shows contours of [O~\textsc{iii}]/UV, which highlights the high-ionization plume to the north of the galaxy nucleus. We find a large-scale velocity gradient from the southeast to the northwest, with an absolute variation of $\sim$1200 km s$^{-1}$. This strong kinematic signature could be a result of bulk rotation, infalling gas, or outflows, depending on the orientation.}
\label{fig:vel}
\end{figure*}

\begin{figure*}[htb]
\centering
\includegraphics[width=0.99\textwidth]{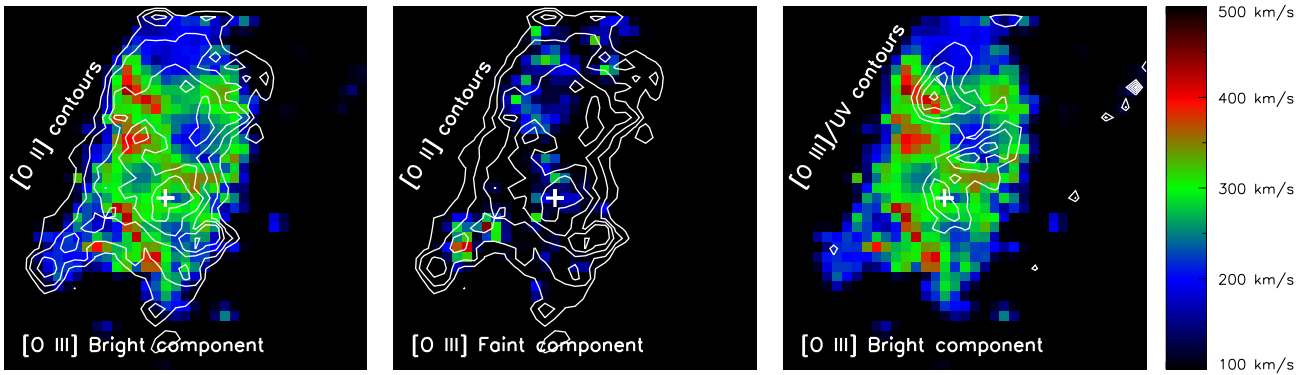}
\caption{Velocity dispersion maps of the warm ionized gas in the core of the Phoenix cluster. The physical scale in all panels is identical to Figure \ref{fig:lines}.  In the central panel we show the velocity dispersion of the second, fainter component in cases where multiple velocity components were detected (see also Fig.\ \ref{fig:doubleline}). In the central and left panels, the white contours represent the [O~\textsc{ii}] emission, while the right-most panel shows contours of [O~\textsc{iii}]/UV, which highlights the high-ionization plume to the north of the galaxy nucleus. We find a large velocity width overall, with an average linewidth of FWHM $\sim$ 700 km s$^{-1}$ (FWHM = 2.355$\sigma$), and very little structure. There is a slight increase in the dispersion northeast of the emission peak, coincident with the high-ionization peak (right-most panel). In general we do not observe a significant correlation between the ionization state and the line width, suggesting that either shocks are not the dominant source of ionization, or some additional process is adding large-scale turbulence to the gas.}
\label{fig:sig}
\end{figure*}

The resulting radial velocity and velocity dispersion maps are shown in Figures \ref{fig:vel} and \ref{fig:sig}, respectively, where all velocities are with respect to $z=0.597$.  In general we observe a relatively smooth velocity gradient from the southeast (+700 km s$^{-1}$) to the northwest (--400 km s$^{-1}$), with the minimum ($v \sim 0$ km s$^{-1}$) occurring near the center of the [O~\textsc{ii}] emission. It is difficult to interpret the physical meaning of these kinematics, as they could result from rotating, infalling, or outflowing gas, observed from different viewing angles.

The velocity width of the [O~\textsc{iii}] emission lines, shown in Figure \ref{fig:sig}, are significantly broadened to the northeast of the emission peak, slightly offset from the direction of the highly-ionized plume shown in Figure \ref{fig:uvrats}. Overall, the velocity dispersion is very high, with a median value of FWHM$_{\hspace{0.5mm}[O~III]} \sim 700$ km s$^{-1}$. This is significantly higher than typical low-z cool core clusters, which have emission-line nebulae with linewidths spanning the range $100 < FWHM < 600$ km s$^{-1}$ \citep[e.g.,][]{mcdonald12a}. 
Given the exceptionally high star formation rate, combined with the presence of a powerful AGN, there are several possible origins for the broad linewidths (e.g., type-II supernovae, radiation-driven winds, AGN-driven winds, etc), so it is perhaps unsurprising that the linewidths are elevated throughout the ISM of the central galaxy.
We find little correlation between the [O~\textsc{iii}] line width and the [O~\textsc{iii}]/H$\beta$ line ratio, with the exception of the high-ionization peak at ``region B'', which has both high He~\textsc{ii}/H$\beta$ and high velocity width ($\sim$1000 km/s). If shocks were the only contributor to the ionization, one would expect these quantities to be correlated across the entire field of view, which we do not observe. The lack of a correlation suggests a much more complex environment, with likely multiple ionization sources and significant turbulence/mixing.

We find a relatively strong velocity gradient along the highly-ionized plume (Figure \ref{fig:vel}, right-most panel), with a projected change in velocity of $\sim$750 km s$^{-1}$ over a distance of only $\sim$10 kpc. This could be indicative of rotation, which would imply an enclosed mass of $\sim$ 10$^{11}$ M$_{\odot}$ in the central $\sim$5~kpc. For comparison, this is an order of magnitude higher than the total mass in the central $\sim$5~kpc of M87, the central galaxy in the Virgo cluster \citep{gebhardt09}. An alternative possible explanation for this strong gradient is a high-velocity outflow in the direction of the highly-ionized plume, centered on the AGN.
This outflow velocity is typical of massive galactic winds \citep{veilleux05}, with the ionized wind in M82 having a deprojected outflow velocity of $\sim$600 km s$^{-1}$ \citep{shopbell98}. The warm, ionized gas appears to reach peak speeds along the highly-ionized plume, with the extended low-ionization material spanning a relatively small range in velocity. The kinematics of this gas may be influenced by an AGN-driven outflow, or general bulk motions of the gas (infalling cool material, rotation, etc.).

\subsection{Cold Molecular Gas Traced by CO(3-2)}





\begin{figure}[htb]
\centering
\includegraphics[width=0.49\textwidth]{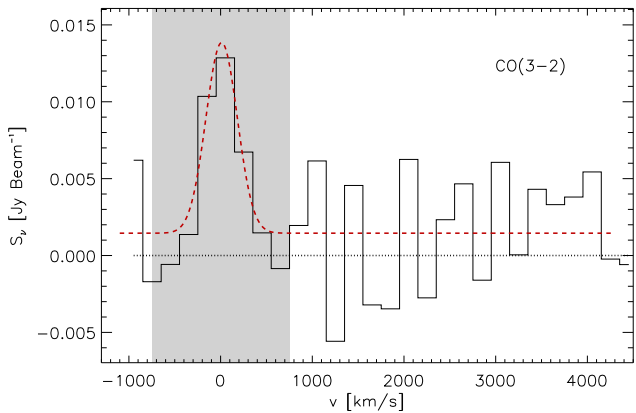}
\caption{CO(3-2) spectrum in the central galaxy of the Phoenix cluster. This spectrum shows evidence of a peak (3.8$\sigma$) at $-150$ km s$^{-1}$ with respect to the central galaxy ($z=0.597$). The velocity range of the [O~\textsc{iii}]-emitting gas (Figure \ref{fig:vel}) is shown in gray, for comparison. The red dashed line shows the fit to this emission, which yields a flux of 5.3 $\pm$ 1.4 Jy km/s and a linewidth of FWHM $\sim$ 400 km s$^{-1}$.}
\label{fig:CO}
\end{figure}

In Figure \ref{fig:CO} we present a 3.8$\sigma$ detection of CO(3-2), in the core of the Phoenix cluster. While this detection significance is low, it is bolstered by the proximity to Phoenix A in both position (1.3$^{\prime\prime}$) and velocity (+11 km s$^{-1}$). We measure a line flux of S$_{CO(3-2)}$ = 5.3 $\pm$ 1.4 Jy km/s. Following \cite{solomon92}, we calculate $L'_{CO(3-2)} = 3.25\times10^7S_{CO(3-2)}\Delta v\nu_{obs}^{-2}D_L^2(1+z)^{-3}$ = 1.1 $\pm$ 0.3 $\times$ 10$^{10}$ K km s$^{-1}$ pc$^2$. 
Assuming a (3-2)/(1-0) ratio of $r_{31} \sim 0.5$, which is typical of dusty starburst galaxies \citep{yao03,iono09,leech10,papadopoulos12} and only slightly lower than that found for Abell~1835 \citep[$r_{31} \sim 0.85$;][]{edge01},we infer a luminosity of $L'_{CO(1-0)} = 2.2 \pm 0.6$ $\times$ 10$^{10}$ K km s$^{-1}$ pc$^2$.

The inferred H$_2$ mass is, of course, very sensitive to our choice of the CO-to-H$_2$ conversion, $\alpha_{CO}$, but the latter is quite uncertain \citep[see][for a review]{bolatto13}. Previous studies of cool core clusters have used the Galactic value of $\alpha_{CO} \sim 4$ M$_{\odot}$ pc$^{-2}$ (K km/s)$^{-1}$ \citep[e.g.,][]{edge01,salome03}, while studies of ULIRGs and other starburst galaxies have found values an order of magnitude lower: $\alpha_{CO}$ $\sim$ 0.6 $\pm$ 0.2 M$_{\odot}$ pc$^{-2}$ (K km/s)$^{-1}$ \citep[e.g.,][]{papadopoulos12, bolatto13}. While we have insufficient information to properly constrain $\alpha_{CO}$, there are several lines of evidence that lead us to adopt the lower value:

\begin{enumerate}
\item \emph{The average velocity dispersion for the 10$^4$\,K gas in Phoenix A is $\sigma \sim$ 350 km s$^{-1}$}. \cite{shetty11} show that the amount of turbulence in a given giant molecular cloud strongly correlates with the value of $\alpha_{CO}$, such that an increase in the turbulent velocities by a factor of 10 can yield a factor of $>$3 decrease in the value of $\alpha_{CO}$. While the amount of turbulence in the HII and H$_2$ are bound to be different, the fact remains that the 10$^4$\,K gas in Phoenix A has 1-2 orders of magnitude more velocity broadening than a typical disk galaxy.
\item \emph{The typical star formation surface density in the central $\sim$10 kpc is $\sim$5 M$_{\odot}$ yr$^{-1}$ kpc$^{-2}$}. This number is typical of starburst galaxies ($\alpha_{CO} \lesssim 1$), and orders of magnitude higher than what is measured in the Galactic disk ($\alpha_{CO} \sim 4$). 
\item \emph{The metallicity of the ICM in the central 50\,kpc is $\sim$1.5Z$_{\odot}$}. Under the assumption that the cooling ICM is the source of the cold gas reservoir, this implies that $\alpha_{CO}$ should be lower than Galactic \citep{bolatto13}.
\item \emph{The dust temperature in Phoenix A is 87\,K}. This is higher than in a typical star-forming galaxy, and is consistent with a lower value of $\alpha_{CO}$, assuming that the cloud temperature and dust temperature are related \citep[$\alpha_{CO} \propto (\sigma T)^{-1}$;][]{bolatto13}.
\end{enumerate}

\noindent{}Based on these arguments, we choose to adopt a lower, starburst-like value for $\alpha_{CO}$. In the interest of being conservative, we will assume $\alpha_{CO} = 1.0$ for Phoenix A, but we will consider the full range of realistic values (0.4--4.0) throughout the discussion. This choice of $\alpha_{CO}$ leads to an estimate of M$_{H_2}$ = 2.2 $\pm$ 0.6 $\times$10$^{10}$ M$_{\odot}$. At a glance, this number may seem low compared to other strong cool core clusters such as Abell~1835 \citep[M$_{H_2}$ = 9.2 $\times$10$^{10}$ M$_{\odot}$;][]{edge01, mcnamara13} and Zw3146 \citep[M$_{H_2}$ = 8.2 $\times$10$^{10}$ M$_{\odot}$;][]{edge01}, which have substantially lower star formation rates than the central galaxy in the Phoenix cluster.  However, if we instead assume a Galactic value of the CO-to-H$_2$ conversion ($\alpha_{CO} = 4$), we arrive at a much higher estimate of the total molecular gas mass: $\sim$8.8 $\times$ 10$^{10}$ M$_{\odot}$.  Assuming that the same value of $\alpha_{CO}$ characterizes all starbursts in cool cores, we conclude that Phoenix, Abell~1835, and Zw3146 all harbor similar, if uncertain, quantities of cold molecular gas at their centers.

At the current rate of star formation \citep[$\sim$800 M$_{\odot}$ yr$^{-1}$;][]{mcdonald13a}, and assuming the starburst value for $\alpha_{CO}$, the current supply of cold molecular gas would be exhausted in less than 30 Myr, provided there is no compensating source replenishing the cold gas reservoir. Given that the typical BCG stellar mass in a cluster the size of Phoenix is $\sim$8$\times$10$^{11}$ M$_{\odot}$ \citep{lidman12}, this starburst may provide $\sim$3\% of the total BCG stellar mass (or more, depending on how long it has been ongoing and whether or not it is being replenished). 

\begin{figure}[htb]
\centering
\includegraphics[width=0.4\textwidth]{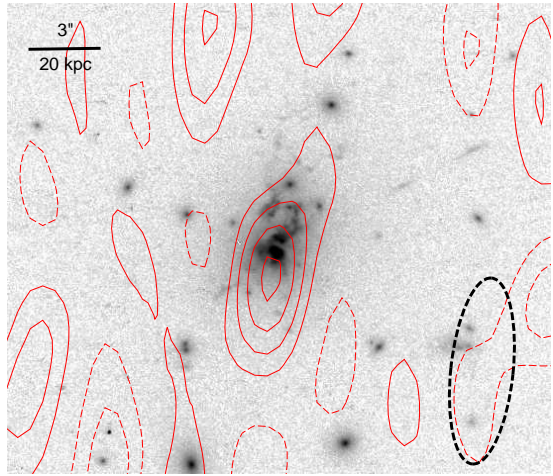}
\caption{CO(3-2) channel map, made by combining the $-$300, $-$100, and $+$100 km s$^{-1}$ channels, overlaid in red contours on the HST F814W image \citep{mcdonald13a}. Solid and dashed contours represent $\pm$1$\sigma$, $\pm$2$\sigma$, etc. The size of the beam, shown in the lower right, is roughly the same size as the CO(3-2) detection, suggesting a lack of detected extended emission. The correspondence in position and radial velocity (Figure \ref{fig:CO}) between the warm ionized gas and the CO(3-2) emission further strengthens the significance of this detection.}
\label{fig:COmap}
\end{figure}

The spatial distribution of the CO(3-2) emission is shown in Figure \ref{fig:COmap}. This map, which is overlaid on an HST F814W image, was generated by combining the $-150$, $+50$, and $+250$ km s$^{-1}$ channels. There appears to be a slight offset ($\sim$1$^{\prime\prime}$, $\sim$7 kpc) between the peak of star-formation and the peak of the CO(3-2) emission, which is larger than our absolute astrometric uncertainty ($\sim$0.2$^{\prime\prime}$), but similar in size to our centroiding uncertainty (\textsc{fwhm}/(\textsc{s/n}) $\sim$ 8$^{\prime\prime}$/4 $\sim$ 2$^{\prime\prime}$). The emission shown in Figure \ref{fig:COmap} may be marginally extended, but not significantly more so than the beam ($7.5^{\prime\prime}\times2.4^{\prime\prime}$). Future ALMA observations with improved spatial resolution will be able to properly quantify any spatial offset and determine the morphology of this cold gas reservoir.


\section{Discussion}
In previous work \citep{mcdonald12c, mcdonald13a} we presented compelling evidence for a massive starburst in the central galaxy of the Phoenix cluster, based largely on the presence of morphologically complex UV continuum emission, combined with an exceptionally high far-IR luminosity. 
The results presented thus far depict a much more complex system, with a highly-ionized nucleus and plume to the north of the cluster core, a massive cold gas reservoir, and substantial turbulent motions in the warm gas. Below, we discuss the implications of these new findings, and attempt to paint a more complete picture of the ongoing processes in this system.

\subsection{The Extreme Nature of the Phoenix Starburst}

\begin{figure}[htb]
\centering
\begin{tabular}{c}
\includegraphics[width=0.46\textwidth]{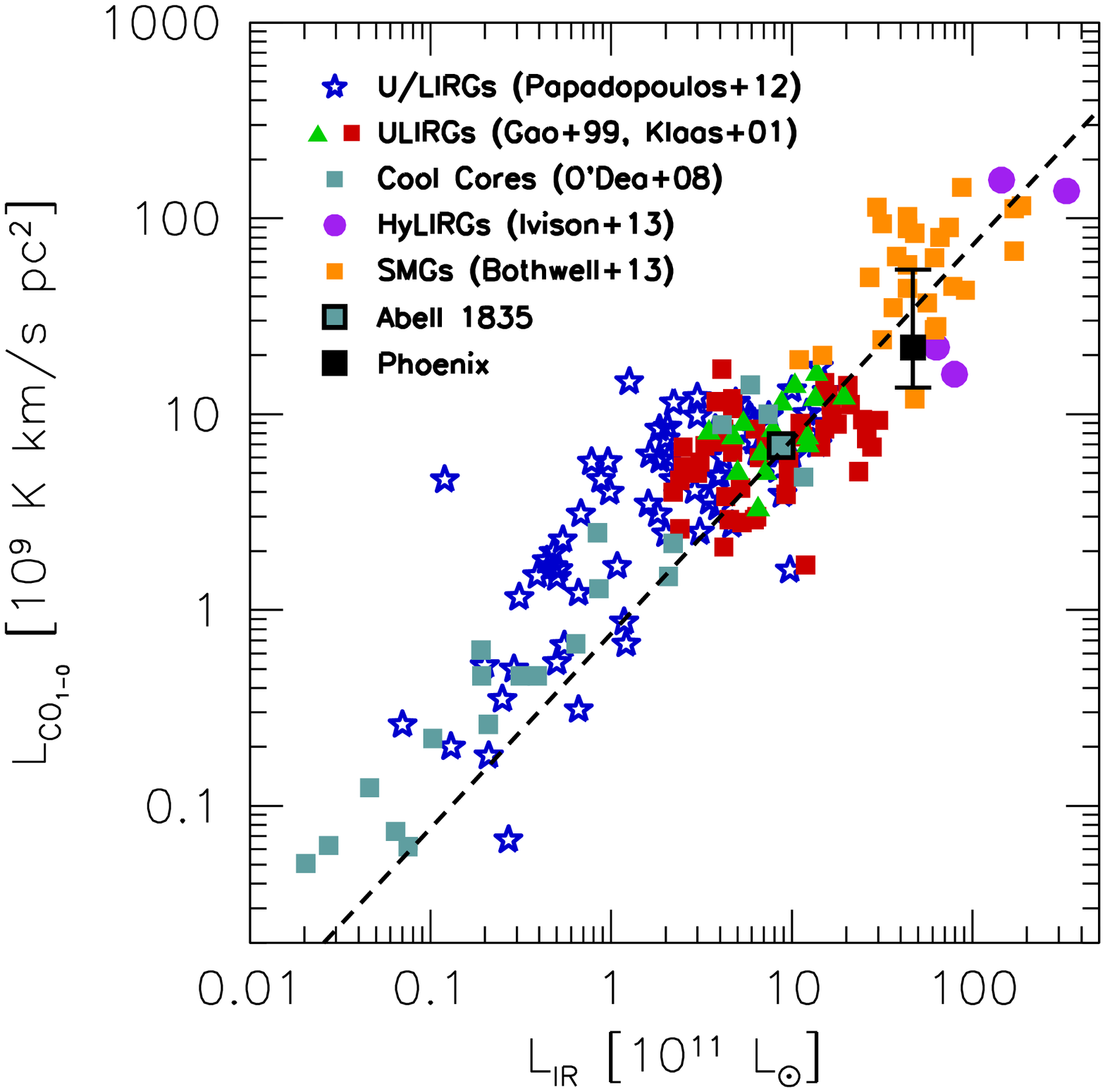} \\
\includegraphics[width=0.46\textwidth]{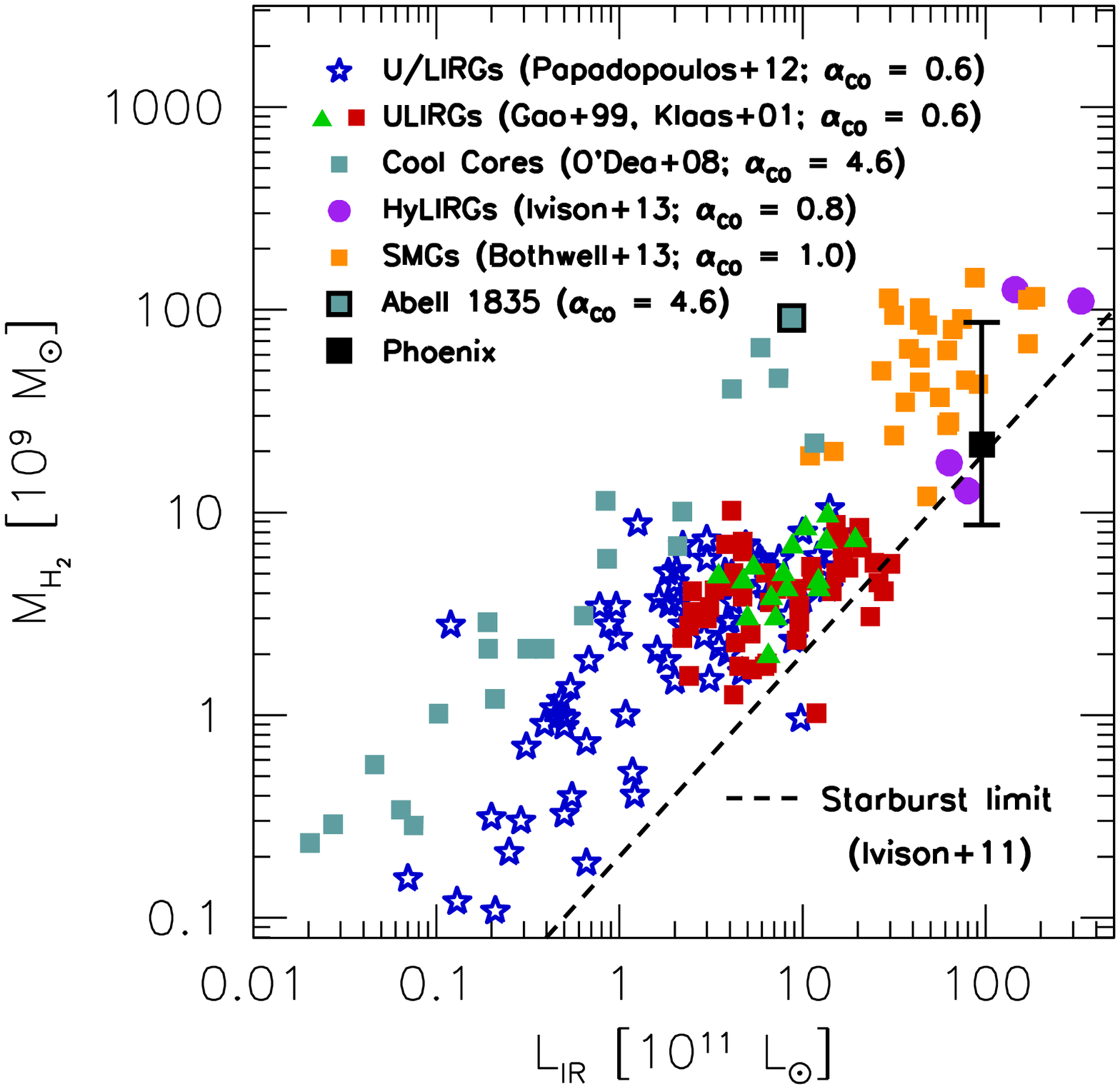} \\
\end{tabular}
\caption{Upper panel: CO(1-0) luminosity as a function of total infrared luminosity for a wide variety of galaxies. Phoenix A, lying in the upper right, has a typical L$_{CO}$/L$_{IR}$ ratio, with absolute values in the same regime as high-z submillimeter galaxies \citep{bothwell13} and hyperluminous infrared galaxies \citep{ivison13}. The error range for L$_{CO_{1-0}}$ represents our uncertainty in the value of $r_{31}$ (see \S3.4). Lower panel: Molecular gas mass as a function of total infrared luminosity. Here, M$_{H_2}$ was taken directly from various publications, which assume a wide range of $\alpha_{CO}$ values from 0.6--4.6. The extreme starburst limit, from \cite{ivison11}, corresponds to the star formation rate at which radiation pressure is sufficient to disperse the cold molecular gas. The fact that the Phoenix cluster lies on this line (assuming a starburst-like $\alpha_{CO}$) suggests that it may be in the midst of quenching its own star formation. The error range for M$_{H_2}$ represents our uncertainty in the intrinsic value of $\alpha_{CO}$ (see \S3.4).
In the upper panel, we have removed the AGN contribution ($\sim$50\%) to the IR luminosity of Phoenix A by assuming a star formation rate of 800 M$_{\odot}$ yr$^{-1}$ \citep[see][]{mcdonald12c}, while in the lower panel we have taken the total IR luminosity, as the AGN is certainly contributing to the amount of radiation pressure.}
\label{fig:lir2lco}
\end{figure}

In \cite{mcdonald13a} we report a star formation rate in Phoenix A of $\sim$800 M$_{\odot}$ yr$^{-1}$. For context, the next most rapidly star-forming brightest cluster galaxy lies at the center of Abell 1835, with a star formation rate of $\sim$150 M$_{\odot}$ yr$^{-1}$ \citep{mcnamara06}. The addition of new CO(3-2) observations provide further insights into just how extreme Phoenix A is. In the upper panel of Figure \ref{fig:lir2lco} we show the correlation between the IR luminosity and the CO(1-0) luminosity, for a variety of systems, including cool core BCGs \citep{odea08}, luminous infrared galaxies \citep[LIRGs;][]{papadopoulos12}, ultraluminous infrared galaxies \citep[ULIRGs;][]{gao99, klaas01}, hyperluminous infrared galaxies \citep[HyLIRGs;][]{ivison13}, and high-redshift submillimeter galaxies \citep[SMGs;][]{bothwell13}. By comparing luminosities rather than derived products such as M$_{H_2}$ and M$_{dust}$, we avoid the large uncertainties in conversion coefficients such as $\alpha_{CO}$. This figure shows a tight correlation over five decades in luminosity, with the central galaxy in the Phoenix cluster having similar IR and CO luminosity to HyLIRGs and SMGs. 

In the lower panel of Figure \ref{fig:lir2lco} we compare the H$_2$ mass to the total infrared luminosity for the same sample. Here we have taken M$_{H_2}$ directly from various authors, meaning that these data span a range of $\alpha_{CO}$ from 0.6--4.6. This plot shows that Phoenix A, with a star formation rate of $\sim$800 M$_{\odot}$ yr$^{-1}$, lies on the ``extreme starburst limit'', assuming a starburst-like value of the CO-to-H$_2$ conversion ($\alpha_{CO}$\,$\lesssim$~1). This limit, corresponding to L$_{IR}$/M$_{H_2}$ = 500 L$_{\odot}$/M$_{\odot}$, represents the IR luminosity at which radiation pressure is able to disperse the cold gas reservoir \citep[see e.g.,][]{thompson05, ivison11}. The fact that Phoenix A lies right on this line implies an exceptionally vigorous starburst, to the point of nearly tearing itself apart. Given that the warm (10$^4$K) gas is more susceptible to the effects of radiation pressure, it is indeed unsurprising that the turbulent gas motions and ionization ratios are so high throughout this central galaxy, as it is likely being repeatedly shocked as it is driven outwards into the ICM by the starburst.

\subsection{A Dust-Obscured AGN}
In \cite{mcdonald12c}, we reported the presence of a dusty AGN in the central galaxy of the Phoenix cluster, based on the presence of a heavily-obscured X-ray point source and an exceptionally high IR luminosity. This AGN was later confirmed to be a type-2 QSO by \cite{ueda13} based on combined Suzaku and Chandra observations. 
This is one of only a small number of type-2 QSO discovered in the cores of cooling flow clusters, with the others being IRAS 09104+4109 \citep[e.g.,][]{osullivan12} and Cygnus-A \citep{djorgovski91}. With these new data, we can further confirm that the reddening peaks on the nucleus (Figure \ref{fig:ebv}), that there are high-ionization emission lines ([Ne~\textsc{iii}], He~\textsc{ii}) coincident with the galaxy center (Figure \ref{fig:lines}), and that the emission line ratios are consistent with the expectation for photoionization by an AGN in the presence of dust \citep[Figure \ref{fig:linerats};][]{groves04b}. These data seem to paint a fairly clear picture of a dust-enshrouded AGN, although it remains unclear how much of an effect it is having on the surrounding ISM and ICM. We address the possible influence of this AGN on the ISM in the next section, and await deeper X-ray observations in order to determine whether it is influencing the larger-scale ICM.

\subsection{A Highly-Ionized AGN/Starburst-Driven Wind?}
In Figure \ref{fig:lines} we show a highly-ionized plume of warm gas extending north from the central galaxy in the Phoenix cluster. This plume, detected at He~\textsc{ii}, has substantially higher [O~\textsc{iii}]/H$\beta$ and [O\textsc{iii}]/UV ratios than the surrounding gas, suggesting a separate ionizing mechanism. In addition, we show in Figure \ref{fig:vel} an $\sim$800 km s$^{-1}$ velocity gradient along the base of this plume. Taken together, this evidence is consistent with the presence of a galactic wind \citep[see e.g.,][]{veilleux05}. We propose that some combination of both mechanical (e.g., AGN jets, SNe ejecta) and radiation pressure from both the AGN and the starburst are acting to drive the cooling material out of the cluster core. 
As this gas interacts with the slower-moving cool ISM/ICM it is shock-excited which, combined with photoionization from the AGN itself, yields high-ionization lines (e.g., [O~\textsc{iii}], He~\textsc{ii}; Fig.\ \ref{fig:lines}), elevated high-ionization line ratios (e.g., [O~\textsc{iii}]/H$\beta$ $>$ 5, He~\textsc{ii}/H$\beta$ $>$ 0.1; Fig.\ \ref{fig:lineratims}), and large velocity dispersions -- all of which are observed in ``region B'' to the northeast of the galaxy center.
The narrow opening angle of this high-velocity plume seems to suggest that the wind is more likely launched by radio jets. The powerful radio source (see Figure \ref{fig:sed}) in Phoenix A is certainly energetic enough to drive such a wind, and there is some evidence that radio jets can drive an ionized outflow in nearby clusters \citep[e.g.][]{werner11, farage12}. However, confirmation of this hypothesis awaits radio observations with higher angular resolution and broader frequency coverage. 

As shown in Figure \ref{fig:linerats}, the range of observed line ratios can be explained by both shocks and AGN photoionization, with stellar photoionization being more important for the low-ionization lines. Our proposed explanation for all of the observations from X-ray to radio is that the Phoenix cluster is undergoing a short-lived phase of rapid cooling, leading to both a massive starburst and rapidly-accreting AGN \citep[$\sim$60~M$_{\odot}$ yr$^{-1}$;][]{mcdonald12c} that are quickly consuming the accumulated cold gas reservoir. Radiation pressure from the starburst and the obscured type-2 QSO \citep{ueda13}, combined with the powerful radio jets, is most likely driving a large-scale wind to the north of the central galaxy, shocking the high-velocity material against the cooling ICM (region B), leading to a plume of high-ionization material. Simultaneously, the dust-obscured AGN is locally photoionizing the nucleus (region A) and likely contributing to the ionization in the highly-ionized plume, while the star formation is providing lower levels of photoionization over the full area of the starburst.

\subsection{The Phoenix Cluster: An Extreme, Yet Normal, Cooling Flow or an Outlier?}

With the addition of spatially resolved optical spectroscopy and deep sub-mm data, a more complete picture of the core of the Phoenix cluster is emerging. 
The spectral energy distribution of Phoenix A, spanning the far-UV through sub-mm (Figure \ref{fig:sed}), is reminiscent of a dusty starburst with an obscured, radio-loud type-2 QSO at its center. The starburst, which we suspect is fueled by the 2700 M$_{\odot}$ yr$^{-1}$ cooling flow \citep{mcdonald12c}, is operating near maximum efficiency (L$_{IR}$/M$_{H_2}$ $\sim$ 500 L$_{\odot}$/M$_{\odot}$), rapidly depleting the cold gas reservoir over timescales of $\lesssim$\,30 Myr while threatening to disperse the molecular gas on even shorter timescales due to the immense radiation pressure. The central AGN, perhaps in combination with radiation pressure from the starburst, appears to be driving an outflow to the north of the central galaxy, with the He~\textsc{ii}/UV ratio peaking $\sim$15~kpc north of the AGN (Figure \ref{fig:uvrats}). 

\begin{figure}[htb]
\centering
\includegraphics[width=0.45\textwidth]{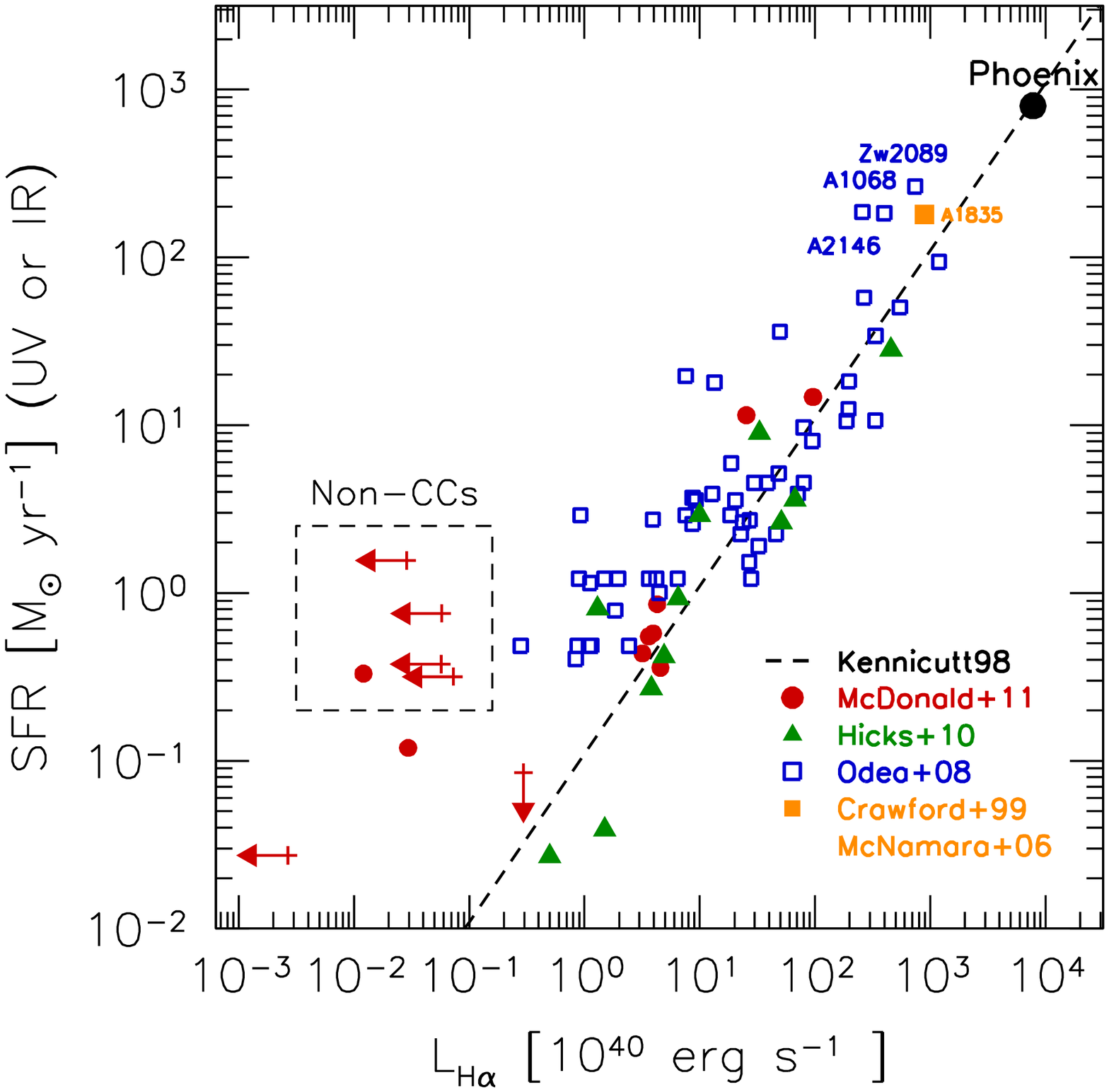}
\caption{Star formation rate (SFR) versus H$\alpha$ luminosity for a sample of cool core BCGs taken from the literature \citep{crawford99,mcnamara06,odea08,hicks10,mcdonald11b}. SFRs are based on far-UV or far-IR continuum. If the emission-line nebulae in cool core clusters were predominantly photoionized, the data should follow the dashed line \citep{kennicutt98}. For the most part, this is the case, suggesting that alternative ionization mechanisms (i.e., shocks, particle heating, etc) are secondary. For Phoenix, we have assumed L$_{H\alpha}$/L$_{H\beta}$ = 2.85. This plot shows that, while extreme in nature, the low-ionization lines in the central galaxy of the Phoenix cluster are most likely the result of photoionization.}
\label{fig:uvha}
\end{figure}

\begin{figure}[h!]
\centering
\includegraphics[width=0.46\textwidth]{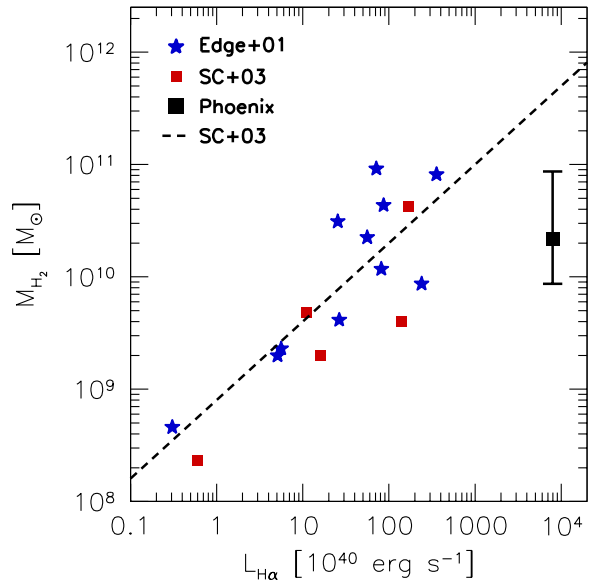} \\
\caption{Molecular gas mass (M$_{H_2}$) as a function of H$\alpha$ luminosity (L$_{H\alpha}$). We show here data from \cite{edge01} and \cite{salome03}, for comparison to the Phoenix cluster. The error range shown for M$_{H_2}$ represents our uncertainty in the CO-to-H$_2$ conversion (see \S3.4). This figure demonstrates that the central galaxy in the Phoenix cluster appears to be slightly H$_2$-poor compared to its low-redshift counterparts.}
\label{fig:edge01a}
\end{figure}

This extreme system is almost certainly short-lived. The molecular gas reservoir, at its current size, cannot sustain such a starburst for more than $\sim$30 Myr. At the same time, winds generated by the AGN and radiation pressure from the starburst may eject the cold gas on even shorter timescales, as evidenced by the high L$_{IR}$/M$_{H_2}$ ratio (which includes only the starburst). However, despite the transient nature of this phenomenon, it is worth addressing whether the Phoenix cluster can be thought of as simply a scaled-up version of a typical cool core, or if it is in an altogether different class of object.

In Figure \ref{fig:uvha} we show the star formation rate (measured from UV or IR continuum) versus the H$\alpha$ luminosity for a sample of cool cores drawn from the literature \citep{crawford99, mcnamara06, odea08, hicks10, mcdonald11b}. The correlation between the H$\alpha$ flux and the UV or IR flux has been used to argue that the complex emission-line nebulae observed in cool core BCGs is primarily ionized by young stars. The fact that these data fall along the relationship for photoionization \citep{kennicutt98} supports this scenario. Assuming H$\alpha$/H$\beta$ = 2.85, we find that Phoenix A lies almost exactly along this line, with L$_{H\alpha}$ $\sim$ 8$\times$10$^{43}$ erg s$^{-1}$. While there is nearly an order of magnitude gap between the next most star-forming system \citep[Abell~1835;][]{mcnamara06}, there appears to be little difference in the ionization source between the central galaxies in systems like Abell~1795 \citep[SFR $\sim$ 10M$_{\odot}$ yr$^{-1}$;][]{mcdonald09}, Abell~1835 \citep[SFR $\sim$ 150 M$_{\odot}$ yr$^{-1}$;][]{mcnamara06}, and the Phoenix cluster (SFR $\sim$ 800 M$_{\odot}$ yr$^{-1}$) -- they all appear to have their low-ionization lines produced by photoionization by young stars.

In Figure \ref{fig:edge01a} we compare the H$\alpha$ luminosity to the molecular gas mass, following \cite{edge01} and \cite{salome03}. We show a range of values for $\alpha_{CO}$ in Phoenix A: from a ULIRG-like value of 0.4 to a Galactic value of 4.0 (see \S3.4). This figure demonstrates that Phoenix A appears to be using its molecular gas more efficiently, with a relatively small cold gas reservoir given the enormous star formation rate and H$\alpha$ luminosity. While our choice of $\alpha_{CO}$ is highly uncertain, even assuming a Galactic value (similar to \cite{edge01} and \cite{salome03}) leads to a value of M$_{H_2}$ for Phoenix A that lies below the extrapolation from nearby systems.


In general, the low-ionization optical ([O~\textsc{ii}], H$\beta$) emission lines in Phoenix A appear to be consistent with a ``scaling-up'' of classical low-redshift systems such as Abell~1795 and Abell~1835. Similar to low-redshift systems, the low-ionization lines appear to be ionized by young stars, while the high-ionization lines are produced by some combination of AGN photoionization and shocks. The primary difference between Phoenix A and a scaled-up, low-redshift BCG is the efficiency with which the cooling flow is converted into stars, which sits at $\sim$30\%, compared to the typical low-z value of a few percent. This is also reflected in the ratio of star formation to cold molecular gas, which is much higher for Phoenix A than for nearby BCGs (regardless of our choice of $\alpha_{CO}$). It appears that this highly-efficient cooling phase must be short-lived, based on the current mass of the cold gas reservoir, unless this reservoir is being replenished. The starburst will likely cease in $\sim$30 Myr, leaving a more typical, highly-inefficient cooling flow. It is currently unclear whether this short-lived, highly-efficient phase of cooling is ubiquitous to cooling flows at early times. If it is, one would expect to see post-starburst signatures in a large fraction of high-redshift clusters, as these should survive for $\sim10^8-10^9$ years.

\section{Summary and Conclusions}
We have presented new optical integral field spectroscopy (Gemini GMOS-IFU) of the warm (10$^4$K), ionized gas and submillimeter spectroscopy (SMA) of the cold, molecular gas in the core of the Phoenix cluster. These new data, combined with the existing multiwavelength observations spanning the X-ray through radio, provide the most complete picture to date of the complex, cooling flow-fed starburst in the core of this cluster. The key results of this study can be summarized as follows:

\begin{itemize}
\item The central galaxy in the Phoenix cluster has the most luminous emission-line nebula of any known cool core cluster, with a total H$\alpha$ luminosity of L$_{H\alpha}$ = 7.6 $\pm$ 0.4 $\times$10$^{43}$ erg s$^{-1}$ (assuming H$\alpha$/H$\beta$ = 2.85).
\item The morphology of the [O~\textsc{ii}] emission agrees well with the high-spatial-resolution far-UV imaging presented in \cite{mcdonald13a}, suggesting that photoionization from young stars is the dominant ionizing mechanism for the low-ionization lines (e.g., [O~\textsc{ii}], H$\beta$).
\item The warm, ionized gas appears to have three distinct phases:
\renewcommand{\theenumi}{\roman{enumi}}%
\begin{enumerate}
\item A highly-ionized nucleus, within which the emission line ratios are consistent with photoionization by an AGN.
\item A high-velocity ($\Delta v \sim 800$ km s$^{-1}$ over $\sim$10 kpc), high-ionization ([O~\textsc{iii}]/H$\beta > 3$) plume to the north of the nucleus. This phase, which is dominated by high-ionization lines ([O~\textsc{iii}], [Ne~\textsc{iii}], He~\textsc{ii}) and characterized by a relative lack of UV continuum emission ([O~\textsc{iii}]/UV $>$ 100), is likely ionized by a combination of AGN photoionization and shocks, and is being driven by some combination of mechanical and radiation pressure from both the starburst and AGN.
\item A large-scale, low-ionization ([O~\textsc{ii}]/H$\beta \lesssim 2$) component. This phase, which is dominated by low-ionization lines ([O~\textsc{ii}], H$\beta$) is likely photoionized by the strong UV component ([O~\textsc{iii}]/UV $\sim$ 10).
\end{enumerate}
\item The highly-ionized plume, which we propose is a high-velocity outflow, appears to be collimated, with a preferred direction to the north of the central cluster galaxy. The He~\textsc{ii}/UV ratio actually peaks $\sim$10--15 kpc from the central galaxy.
\item The velocity dispersion in the warm gas is high ($\sigma_v \gtrsim 200$ km s$^{-1}$) throughout the central galaxy, suggesting a very turbulent environment, consistent with the vigorous starbursts observed in most ULIRGs. The velocity dispersion peaks to the north of the nucleus, at the location of the purported highly-ionized outflow.
\item We measure a cold molecular gas mass of M$_{H_2}$ = 2.2 $\pm$ 0.6 $\times$10$^{10}$ M$_{\odot}$, assuming a CO-to-H$_2$ conversion of $\alpha_{CO} = 1.0$. Our choice of $\alpha_{CO}$ was motivated by the extreme physical environment in Phoenix A, but the cold gas mass could range from 0.9--8.8 $\times$10$^{10}$ M$_{\odot}$ for a realistic range of $\alpha_{CO}$ values. Phoenix A is similar in CO luminosity to other extreme cooling flows (e.g., Abell~1835, Zw3146). 
\item Given the molecular gas mass of M$_{H_2}$ = 2.2 $\times$ 10$^{10}$ M$_{\odot}$, the starburst will exhaust its fuel supply in $\sim$30 Myr, unless it is being replenished on shorter timescales. If every cluster underwent such a short-lived phase of rapid cooling once in its lifetime, then the probability of observing a second Phoenix-like cluster is only $\sim$0.3\%.
\item Phoenix A appears to be undergoing more efficient star formation than typical BCGs in cool core clusters, with a relatively high ratio of L$_{H\alpha}$/M$_{H_2}$ compared to local BCGs.
\item Assuming $\alpha_{CO}$ = 1.0 M$_{\odot}$ pc$^{-2}$ (K km/s)$^{-1}$ and $r_{31}=0.5$, the measured L$_{IR}$/M$_{H_2}$ ratio for the Phoenix cluster is 440 L$_{\odot}$/M$_{\odot}$, which is consistent with the starburst limit of 500 L$_{\odot}$/M$_{\odot}$ \citep{ivison11}. At this level of star formation, there is sufficient radiation pressure to disperse the cold molecular gas, quenching star formation. 
\end{itemize}

The combination of the the strong shock signatures throughout the central galaxy and the high L$_{IR}$/M$_{H_2}$ ratio suggests that the starburst in Phoenix A may be in the midst of destroying its own fuel supply. Even if this were not the case, the cold gas would be exhausted in $\sim$30 Myr at the current star formation rate, implying that this phase of highly-efficient star formation is likely very short lived. As such, we would not expect to observe a substantial number of clusters undergoing a similar process. If such a phenomenon is common in cool cores, however, we would expect to observe a substantial number of high-redshift post-starburst BCGs, as these signatures can survive for 10$^8$--10$^9$ years -- nearly two orders of magnitude longer. These signatures are readily observable with current instrumentation and cluster samples, and would provide new constraints on how common this extreme starburst phase is in the evolution of cooling flows.

\section*{Acknowledgements} 
The authors are grateful to the Directors of the Gemini Telescope and the Submillimeter Array, who graciously provided the observing time for this program, and to Ruta Kale, who provided access to the ATCA data.
M. M. acknowledges support provided by NASA through a Hubble Fellowship grant from STScI. 
M.S. and A.C.E acknowledges support from STFC grant ST/I001573/1. 
S.V. acknowledges support from a Senior NPP Award held at NASA-GSFC and from the Humboldt Foundation to fund a long-term visit at MPE in 2012. 
BAB is supported by the National Science Foundation through grant ANT-0638937, with partial support provided by NSF grant PHY-1125897, the Kavli Foundation, and Chandra Award Number 13800883 issued by the CXC.
M.H. acknowledges support from an STFC studentship.
M.B.B. acknowledges support from NSF through Grant AST-1009012.


\end{document}